\documentclass[twocolumn,times]{aastex61}
\hypersetup{linkcolor=red,citecolor=green,filecolor=cyan,urlcolor=magenta}
\received{}
\revised{}
\accepted{}
\shorttitle{Absolute Distances to SNe Ia}
\shortauthors{Vink\'o et al.}
\submitjournal{PASP}
\begin{document}
\title{Absolute Distances to Nearby Type Ia Supernovae via Light Curve Fitting Methods} 
\correspondingauthor{J. Vink\'o}
\email{vinko@konkoly.hu}

\author{J. Vink\'o}
\affil{Konkoly Observatory of the Hungarian Academy of Sciences,
Konkoly-Thege ut 15-17, Budapest, 1121, Hungary}
\affil{Department of Optics and Quantum Electronics, University of Szeged, 
Dom ter 9, Szeged, 6720, Hungary}
\affil{Department of Astronomy, University of Texas at Austin,
2515 Speedway, Austin, TX, USA}

\author{A. Ordasi}
\affil{Konkoly Observatory of the Hungarian Academy of Sciences,
Konkoly-Thege ut 15-17, Budapest, 1121, Hungary}

\author{T. Szalai}
\affil{Department of Optics and Quantum Electronics, University of Szeged, 
Dom ter 9, Szeged, 6720, Hungary}

\author{K. S\'arneczky}
\affil{Konkoly Observatory of the Hungarian Academy of Sciences,
Konkoly-Thege ut 15-17, Budapest, 1121, Hungary}

\author{E. B\'anyai}
\affil{Konkoly Observatory of the Hungarian Academy of Sciences,
Konkoly-Thege ut 15-17, Budapest, 1121, Hungary}

\author{I.~B. B{\'i}r\'o}
\affil{Baja Observatory of the University of Szeged,
Szegedi ut KT 766, Baja, 6500, Hungary}

\author{T. Borkovits}
\affil{Baja Observatory of the University of Szeged,
Szegedi ut KT 766, Baja, 6500, Hungary}

\affil{Konkoly Observatory of the Hungarian Academy of Sciences,
Konkoly-Thege ut 15-17, Budapest, 1121, Hungary}
\affil{Department of Astronomy, E\"otv\"os Lor\'and University,
Pazmany setany 1/A, Budapest, Hungary}
\author{T. Heged\"us}
\affil{Baja Observatory of the University of Szeged,
Szegedi ut KT 766, Baja, 6500, Hungary}

\author{G. Hodos\'an}
\affil{Konkoly Observatory of the Hungarian Academy of Sciences,
Konkoly-Thege ut 15-17, Budapest, 1121, Hungary}
\affil{Department of Astronomy, E\"otv\"os Lor\'and University,
Pazmany setany 1/A, Budapest, Hungary}
\affil{Centre for Exoplanet Science, School of Physics and Astronomy 
University of St Andrews, St Andrews KY16 9SS, UK}

\author{J. Kelemen}
\affil{Konkoly Observatory of the Hungarian Academy of Sciences, 
Konkoly-Thege ut 15-17, Budapest, 1121, Hungary}

\author{P. Klagyivik}
\affil{Department of Astronomy, E\"otv\"os Lor\'and University,
Pazmany setany 1/A, Budapest, Hungary}
\affil{Konkoly Observatory of the Hungarian Academy of Sciences, 
Konkoly-Thege ut 15-17, Budapest, 1121, Hungary}
\affil{Instituto de Astrofísica de Canarias, C. Vía Láctea S/N, 
38205 La Laguna, Tenerife, Spain}
\affil{Universidad de La Laguna, Dept. de Astrofísica, 
38206 La Laguna, Tenerife, Spain}

\author{L. Kriskovics}
\affil{Konkoly Observatory of the Hungarian Academy of Sciences,
Konkoly-Thege ut 15-17, Budapest, 1121, Hungary}

\author{E. Kun}
\affil{Department of Experimental Physics, University of Szeged,
Dom ter 9, Szeged, 6720, Hungary}

\author{G.~H. Marion}
\affil{Department of Astronomy, University of Texas at Austin, 
2515 Speedway, Austin, TX, USA}

\author{G. Marschalk\'o}
\affil{Department of Astronomy, E\"otv\"os Lor\'and University,
Pazmany setany 1/A, Budapest, Hungary}

\author{L. Moln\'ar}
\affil{Konkoly Observatory of the Hungarian Academy of Sciences, 
Konkoly-Thege ut 15-17, Budapest, 1121, Hungary}

\author{A.~P. Nagy}
\affil{Department of Optics and Quantum Electronics, University of Szeged,
Dom ter 9, Szeged, 6720, Hungary}

\author{A. P\'al}
\affil{Konkoly Observatory of the Hungarian Academy of Sciences,
Konkoly-Thege ut 15-17, Budapest, 1121, Hungary}

\author{J.~M. Silverman}
\affil{Department of Astronomy, University of Texas at Austin, 
2515 Speedway, Austin, TX, USA}
\affil{Samba TV, 123 Townsend St., San Francisco, CA, USA}

\author{R. Szak\'ats}
\affil{Konkoly Observatory of the Hungarian Academy of Sciences,
Konkoly-Thege ut 15-17, Budapest, 1121, Hungary}

\author{E. Szegedi-Elek}
\affil{Konkoly Observatory of the Hungarian Academy of Sciences,
Konkoly-Thege ut 15-17, Budapest, 1121, Hungary}

\author{P. Sz\'ekely}
\affil{Department of Experimental Physics, University of Szeged,
Dom ter 9, Szeged, 6720, Hungary}

\author{A. Szing}
\affil{Baja Observatory of the University of Szeged,
Szegedi ut KT 766, Baja, 6500, Hungary}

\author{K. Vida}
\affil{Konkoly Observatory of the Hungarian Academy of Sciences,
Konkoly-Thege ut 15-17, Budapest, 1121, Hungary}

\author{J.~C. Wheeler}
\affil{Department of Astronomy, University of Texas at Austin, 
2515 Speedway, Austin, TX, USA}

%\date{}

\begin{abstract}
We present a comparative study of absolute distances to a sample of very nearby,
bright Type Ia supernovae (SNe) derived from high cadence, high signal-to-noise, multi-band photometric data. Our sample
consists of four SNe: 2012cg, 2012ht, 2013dy and 2014J. We present new homogeneous, high-cadence photometric data
in Johnson-Cousins $BVRI$ and Sloan $g'r'i'z'$ bands taken from two sites (Piszkesteto and Baja, Hungary), and  
the light curves are analyzed with publicly available light curve fitters (MLCS2k2, SNooPy2 and SALT2.4).
When comparing the best-fit parameters provided by the different codes, it is found that the distance 
moduli of moderately-reddened SNe Ia agree within $\lesssim 0.2$ mag, and the agreement is even better
($\lesssim 0.1$ mag) for the highest signal-to-noise $BVRI$ data. 
For the highly-reddened SN~2014J the dispersion of the inferred distance moduli is slightly higher. 
These SN-based distances are in good agreement with the Cepheid distances to their host galaxies.
We conclude 
that the current state-of-the-art light curve fitters for Type Ia SNe can provide consistent absolute 
distance moduli having less than $\sim 0.1$ -- 0.2 mag uncertainty for nearby SNe. Still, there is room for 
future improvements to reach the desired $\sim 0.05$ mag accuracy in the absolute distance modulus. 
\end{abstract}

%\titlerunning{Light curve fits of bright SNe Ia}
%\authorrunning{Vink\'o et al.}
%\maketitle 

\section{Introduction}\label{intro}

Getting reliable {\it absolute} distances is of immense importance in observational astrophysics. 
Supernovae (SNe) in particular play a central role in establishing the extragalactic distance ladder. 
Distances to Type Ia SNe are essential data for studying the expansion of
the Universe \citep{riess98, per99, astier06, riess07, wood07, kessler09, guy10, conley11, betou14, rest14, scolnic14}. 
SNe Ia are also especially important objects for measuring the Hubble-parameter $H_0$ 
\citep{riess11, riess16, dhawan17}, and they play a key role in testing current cosmological 
models \citep{benitez13, betou14}. 
Their importance has even been strengthened since the release of the cosmological 
parameters from the {\it Planck} mission \citep{planck14, planck15}, 
which turned out to be slightly in tension with the current implementation of the 
SN Ia distance measurements \citep[see][and references therein]{riess16}.

It must be emphasized that the majority of cosmological studies use {\it relative} distances to moderate- 
and high-redshift SNe Ia to derive the cosmological parameters ($\Omega_m$, $\Omega_\Lambda$, $w$, etc). 
From this point of view there is no need for having absolute distances, because the 
relative distances between different SNe/galaxies can be obtained with much better accuracy, 
especially if the galaxy is in the Hubble-flow ($z \gtrsim 0.1$). 

On the other hand, having accurate absolute distances to the nearby galaxies that
are not part of the Hubble-flow is very important an astrophysical point of view. 
Getting reliable estimates for the physical parameters of such galaxies and the objects 
within them is possible only if we have reliable absolute distances on extragalactic scales. 

Thus, investigating the nearest SNe Ia (within $z \lesssim 0.01$) can provide valuable information 
for various reasons. 
For example, distances to their host galaxies can be relatively easily determined by several methods, 
thus, the SN-based distances can be compared directly to those derived independently 
by using other types of objects and/or methods. Such very nearby galaxies might also serve as ``anchors'' in the cosmic distance ladder 
\citep[e.g. NGC~4258, see][]{riess11, riess16}, which play an essential role in measuring $H_0$.

\begin{table*}[h!]
%\centering
\begin{center}
\caption{Basic data for the studied SNe} \label{tab-basic}
\begin{tabular}{lccccccccc}
\hline
\hline
SN & Discovery date & T($B_{max}$) & $\Delta m_{15}(B)$ & Host & $z_{host}$\tablenotemark{a} & $D_{host}$\tablenotemark{b} &$E(B-V)_{MW}$\tablenotemark{c} & $\log M_{*}$\tablenotemark{d} & References \\
 & & (MJD) & (mag) & & & (Mpc) & (mag) & ($M_\odot$) & \\
\hline
\object{SN~2012cg} & 2012-05-17 & 56080.0 & 0.98 & NGC~4424 & 0.001458 & 16.4 & 0.018 & 9.4 & 1,9,10\\
\object{SN~2012ht} & 2012-12-18 & 56295.6 & 1.27 & NGC~3447 & 0.003559 & 24.1 & 0.026 & 9.3 & 2,3,9,11\\
\object{SN~2013dy} & 2013-07-10 & 56501.1 & 0.96 & NGC~7250 & 0.003889 & 20.0 & 0.135 & 9.2 & 4,5,9,12\\
\object{SN~2014J} & 2014-01-21 & 56689.7 & 1.03 & M82 & 0.000677 & 3.9 & 0.140 & 10.5 & 6,7,8,13\\
\hline
\hline
\end{tabular}
\end{center}
\tablenotetext{a}{Host galaxy redshift, adopted from NED}
\tablenotetext{b}{Cepheid distances from \cite{riess16}; mean redshift-independent distance from NED for SN~2014J}
\tablenotetext{c}{Milky Way reddening based on IRAS/DIRBE maps \citep{sf11}}
\tablenotetext{d}{Host galaxy stellar mass based on SED fitting with Z-PEG \citep{leborgne02}}
\tablecomments{
 References: (1):\citet{silver12}; (2):\citet{yusa12}; (3):\citet{yama14}; (4):\citet{zheng13}; (5):\citet{pan15};
 (6):\citet{foss14}; (7):\citet{zheng14}; (8):\citet{marion15} ; (9):\citet{riess16}; (10):\citet{cortes06} ; 
 (11):\citet{mazzei17}; (12):\citet{pan15}; (13):\citet{dale07} }
\end{table*}

The popularity of SNe~Ia as extragalactic distance estimators is mostly due to the fact that their 
absolute distances can be derived via fitting light curves (LCs) of ``normal" Ia events. The LCs of
such events obey the empirical 
Phillips-relation, i.e. intrinsically brighter SNe have more slowly declining LCs in the optical bands
\citep{psk77, phil93}. Even though nowadays SNe Ia seem to be even better standard candles in the 
near-infrared (NIR) regime than in the optical \citep{friedman15,shariff16,weyant17}, obtaining rest-frame
NIR LCs for SNe Ia, except for the nearest and brightest ones, can be challenging. Thus, photometric 
data taken in rest-frame optical bands can still provide valuable information regarding 
distance measurements on the extragalactic scale. 

One of the main motivations of the present paper is to get absolute distances to some of the nearest and brightest
recent SNe Ia by fitting homogeneous, high-cadence, high S/N photometric data with public, widely-used
LC-fitting codes. We selected four nearby Type Ia SNe for this project: 2012cg, 2012ht, 2013dy and 2014J. All of 
them occured in the local Universe, and they were discovered relatively early (more than
1 week before B-band maximum). We have obtained new, densely sampled photometric measurements for each of them 
in various optical bands, which resulted in light curves extending from pre-maximum epochs up to the end of 
the photospheric phase. 
These objects, along with SN~2011fe, belong to the 10 brightest SNe Ia in last decade that were 
accessible from the northern hemisphere. 
However, unlike SN~2011fe, all of them were significantly reddened by dust either in the Milky Way or in 
their hosts, which enabled us to test the performance of the LC-fitters in case of 
reddened SNe. In addition, their very low redshift ($z < 0.01$) eliminated the necessity for K-correction, 
which could be another possible cause for systematic errors when comparing photometry of SNe having 
significantly different redshifts \citep{saunders15}. 
It is also important to note that three out of four SNe in our sample have Cepheid-based distances
obtained by {\it HST}/WFC3, and they were recently used in calibrating $H_0$ with an unprecedented
2.4 percent accuracy \citep{riess16}. 

The basic parameters for the program SNe are collected in Table~\ref{tab-basic}.

In the following we briefly describe the observations (Section~\ref{obs}) and the LC-fitting codes applied (Section~\ref{ana}).
Section~\ref{res} presents the results from the LC fitting, which are discussed further in Section~\ref{disc}. 
Section~\ref{sum} summarizes the main results and conclusions. 

\section{Observations}\label{obs}

Photometry of the target SNe have been carried out at two sites, located $\sim 200$ km apart: at the 
Piszk\'estet{\H o} station of Konkoly Observatory, Hungary, and at Baja Observatory of the University
of Szeged, Hungary. At Konkoly the data were taken with the 0.6m Schmidt telescope through Bessell
$BVRI$ filters. At Baja the observations were carried out with the 0.5m BART telescope equipped
with Sloan $g'r'i'z'$ filters. See \citet{vinko12} for more details on these two instruments.

All data have been reduced using standard {\it IRAF}\footnote{IRAF is distributed by the National Optical Astronomy
Observatories, which are operated by the Association of Universities for Research in Astronomy, Inc., under cooperative agreement with the National Science Foundation.} routines. Transformation to the standard
photometric systems (Johnson-Cousins/Vega and Sloan/AB for the $BVRI$ and $g'r'i'z'$ data, respectively)
was computed using catalogued Sloan-photometry for local tertiary standard stars. In the fields of 
SN~2012cg and SN~2012ht the $BVRI$ magnitudes for the local standards were calculated from their catalogued $g'r'i'z'$ magnitudes using the calibration given by \citet{jordi06}. For SN~2013dy and 2014J the estimated 
$BVRI$ magnitudes obtained this way were cross-checked by observing Landolt standard fields on 
a photometric night and re-calibrating the local standards using the zero-points from the Landolt standards. 
Finally, all $BVRI$ photometry were cross-compared to the Pan-STARRS (PS1) magnitudes
\footnote{http://archive.stsci.edu/panstarrs/search.php} of the local standards stars, and small 
($\lesssim 0.1$ mag) shifts were applied when needed to bring all the photometry to the same zero-point.
The final $BVRI$ magnitudes for the local comparison stars are shown in the Appendix
(Table \ref{tab-12cg-std}, \ref{tab-12ht-std}, \ref{tab-13dy-std} and \ref{tab-14J-std}). 

Photometry of the SNe was obtained via PSF-fitting using DAOPHOT. 
The resulting instrumental magnitudes were transformed 
to the standard systems by applying linear color terms, 
and the zero points of the transformation are tied to the magnitudes of
the local comparison stars.

The final LCs were compared with other published, independent photometry for each SNe, except for SN~2012ht, where
no other available photometry was found. We used the data given by \citet{marion15b}, \citet{pan15} and 
\citet{marion15} for SN~2012cg, 2013dy and 2014J, respectively. A $\sim 0.1$ mag systematic difference was
identified between the $B$-band LCs of the heavily-reddened SN~2014J, which was corrected by shifting 
our data to match those published by \citet{marion15}. No such systematic offsets between our 
data and those from others were found for the remaining three SNe.
The final photometric data can be found in the Appendix, in Tables~\ref{tab-12cg-phot}, \ref{tab-12ht-phot}, 
\ref{tab-13dy-phot1}, \ref{tab-13dy-phot2} and \ref{tab-14J-phot}. 

\section{Analysis}\label{ana}
 
We applied three SN Ia LC-fitters for this study: 
MLCS2k2\footnote{\tt http://www.physics.rutgers.edu/\~{}saurabh/mlcs2k2/}
\citep{riess98, jha99, jrk07}, 
SALT2\footnote{\tt http://supernovae.in2p3.fr/salt/doku.php} 
version 2.4 \citep{guy07, guy10, betou14}
and SNooPy2\footnote{\tt http://csp.obs.carnegiescience.edu/data/snpy/snpy} 
\citep{burns11, burns14}. All these, in principle, rely on 
the Phillips-relation but each code uses different parametrization for fitting
the light curve shape and each has different sets of calibrating SNe (``training sets"). 
There are also other implementations, like SiFTO \citep{conley08} or BayeSN \citep{mandel11},
but the first three listed above have been used most frequently in the literature.

\citet{conley08} categorized such codes as either ``pure LC-fitters" or ``distance calculators".
Distance calculators can provide the true absolute distance as a fitting parameter, but they
require a training set of SNe having independently obtained absolute distances. Since building
such a training set is non-trivial, the calibration of such codes is usually based on a
relatively small number of objects. On the contrary, LC-fitters can predict only relative 
distances, but they can be calibrated using a much larger sample of objects having much more
accurate relative distances. Regarding the three codes applied in our study, MLCS2k2 and
SNooPy2 are distance calculators, while SALT2 is an LC-fitter,
even though it is possible to derive absolute distances from the SALT2 fitting parameters, 
if needed. 

MLCS2k2 \citep{jrk07} uses the following SN~Ia LC model:
\begin{equation}
m_x (\varphi) ~=~ M_x^0 + \mu_0 + \eta_x A_V^0 + P_x \Delta + Q_x \Delta^2 ,
\label{eq2}
\end{equation}
where $\varphi = t-T_{max}$ is the SN phase in days, $T_{max}$ is the moment of maximum light in the $B$-band,
$m_x$ is the observed magnitude in the $x$-band ($x = B,V,R,I$), $M_x^0 (\varphi)$ is the fiducial 
SN~Ia absolute LC in the same band, $\mu_0$ is the true (reddening-free) SN distance modulus, 
$\eta_x = \zeta_x (\alpha_x + \beta_x/R_V)$ gives the time-dependent
interstellar reddening, $R_V$ and $A_V^0$ are the ratio of total-to-selective absorption and
$V$-band extinction at maximum light, respectively, $\Delta$ is the main LC parameter, and
$P_x (\varphi)$ and $Q_x (\varphi)$ are tabulated functions of the SN phase (``LC-vectors"). 
The absolute magnitudes
of SNe in MLCS2k2 have been calibrated using relative distances of more than 150 SNe in the
Hubble-flow assuming $H_0 = 65$ km s$^{-1}$ Mpc$^{-1}$, but later they were tied to Cepheid
distances of a smaller sample of SNe Ia host galaxies \citep{riess05}. 

Contrary to MLCS2k2, SALT2 models the whole spectral energy distribution (SED) of a SN Ia as
\begin{equation}
F_\lambda (\varphi) ~=~ x_0 \cdot [ M_0(\varphi,\lambda) + x_1 M_1(\varphi,\lambda) ] \exp[C \cdot C_L(\lambda)],
\label{eq3}
\end{equation}
where $F_\lambda (\varphi)$ is the phase-dependent
rest-frame flux density, $M_0(\varphi,\lambda)$, $M_1(\varphi,\lambda)$ and $C_L(\lambda)$ are 
the SALT2 trained vectors. The free parameters $x_0$, $x_1$ and $C$ are the normalization- ,
stretch- and color parameters, respectively. 

Being an LC-fitter, SALT2 does not contain the distance as a fitting parameter. Instead, the
distance modulus can be calculated from the following equation \citep{conley11, betou14, scolnic16}:
\begin{equation}
\mu_0 ~=~ m^*_B - M_B + \alpha x_1 - \beta C.
\label{eq4}
\end{equation}
For the nuisance parameters $\alpha$, $\beta$ and $M_B$ we adopted the calibration by \citet{betou14}:
$M_B = -19.17 \pm 0.038$, $\alpha = 0.141 \pm 0.006$, $\beta = 3.099 \pm 0.075$, and derived
the distance moduli from the fitting parameters via Monte-Carlo simulations. 

One of the great advantages of SALT2 is that it can be relatively easily applied to data taken
in practically any photometric system provided the filter transmission functions are loaded into the code.  
Because of this and many other reasons, SALT2 became very popular recently, and it was used in most
papers dealing with SN Ia light curves 
\citep[e.g.][]{betou14,mosher14,scolnic14,rest14,saunders15,walker15,riess16,zhang17}. 
We utilized the built-in ``Landolt-Bessell" and ``SDSS" filter sets for fitting our $BVRI$
and $g'r'i'z'$ data, respectively.

\begin{figure}
\centering
\resizebox{8cm}{!}{\includegraphics{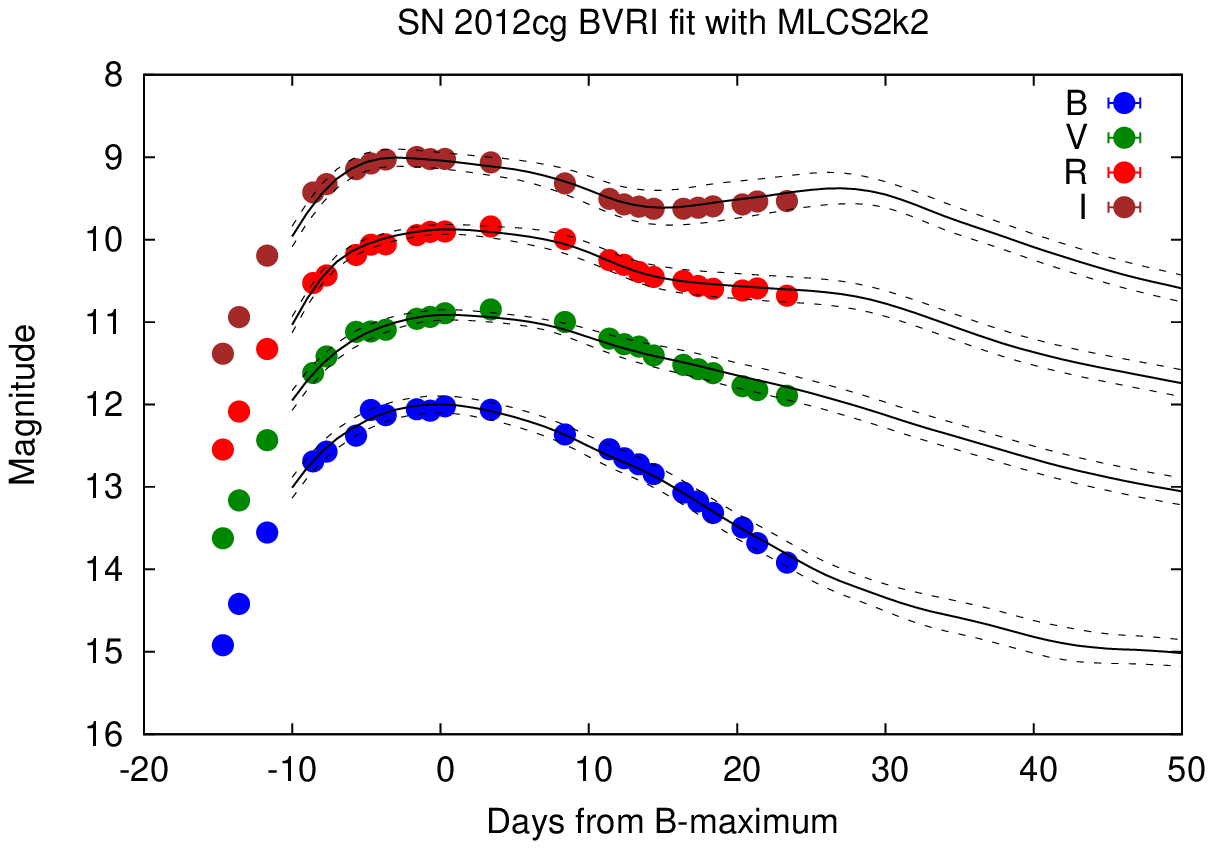}}\\
\resizebox{8cm}{!}{\includegraphics{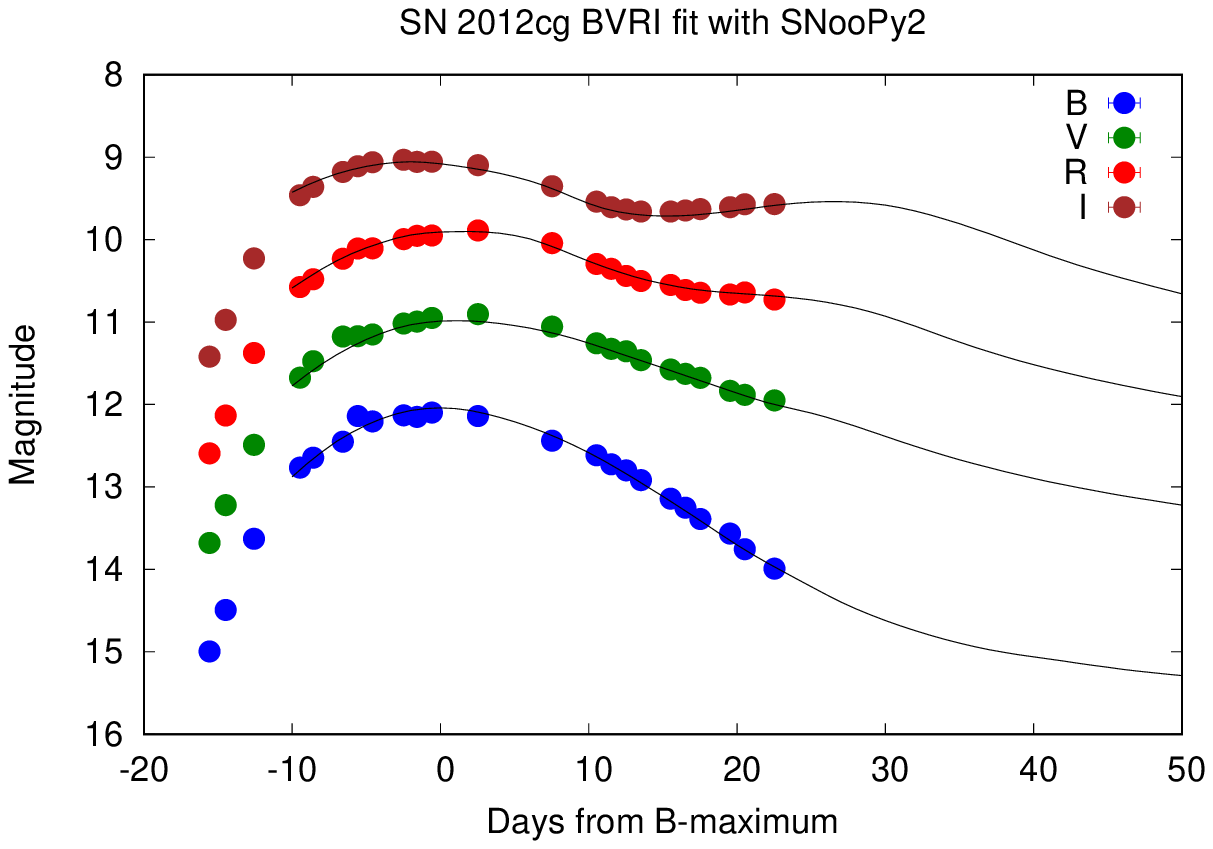}}
\resizebox{8cm}{!}{\includegraphics{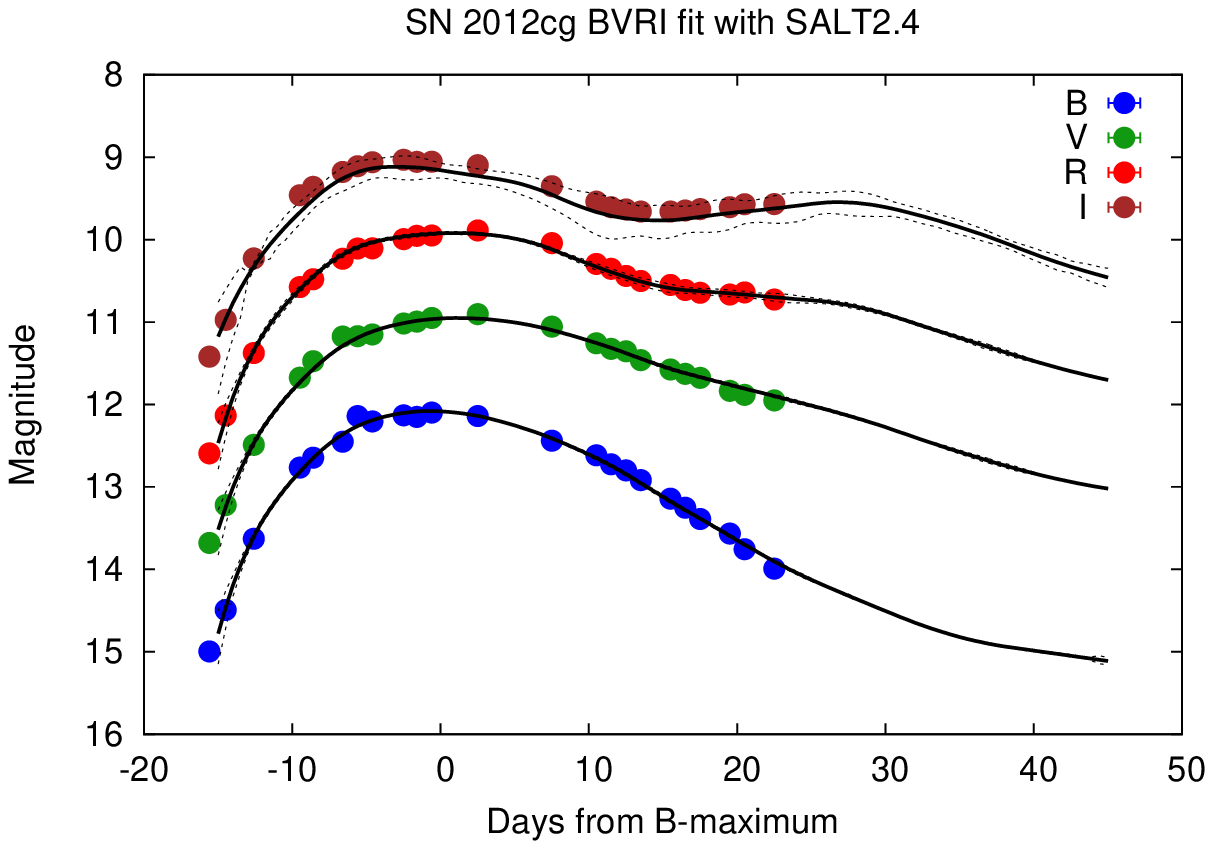}}
\caption{The fitting of SN~2012cg light curves, after correcting for Milky Way extinction.
Top: MLCS2k2 templates; middle: SNooPy2 templates; bottom: SALT2.4 templates. Dashed and dotted 
lines represent the template uncertainties for MLCS2k2 and SALT2.4, respectively.}
\label{fig-12cg}
\end{figure}

While applying SNooPy2, we adopted the default ``EBV-model" as a proxy for a SN Ia LC:
\begin{eqnarray}
m_X (\varphi) ~=~ T_Y + M_Y + \mu_0 + K_{XY} + \nonumber \\
R_X E(B-V)_{MW} + R_Y E(B-V)_{host} , 
\label{eq:ebvmod}
\end{eqnarray}
where $X,Y$ denote the filter of the observed data and the template light curve, respectively, 
$m_X(\varphi)$ is the observed LC in filter $X$, 
$T_Y(\varphi,{\Delta}m_{15})$ is the template LC as a function of time, 
${\Delta}m_{15}$ is the generalized decline-rate parameter associated with the ${\Delta}m_{15}(B)$ parameter by \citet{phil93}, 
$M_Y({\Delta}m_{15})$ is the absolute magnitude of the SN in filter $Y$ as a function of ${\Delta}m_{15}$, 
$\mu_0$ is the reddening-free distance modulus in magnitudes, $E(B-V)$ is the color excess due to interstellar 
extinction either in the Milky Way (``MW"), or in the host galaxy, $R_{X,Y}$ are the reddening slopes in filter
$X$ or $Y$ and $K_{XY}(t,z)$ is the cross-band K-correction that matches the observed broad-band magnitudes of a
redshifted ($z \gtrsim 0.01$) SN taken with filter $X$ to a template SN LC in filter $Y$. Since all our SNe had very low redshift ($z < 0.01$) we always set $X = Y$ and neglected the K-corrections, which greatly simplified the
analysis of those objects.

SNooPy2 offers two sets of templates which cover different filter bands. We utilized the Prieto-templates 
\citep{prieto06} for fitting the $BVRI$ LCs, while for the $g'r'i'z'$ data we selected the 
built-in CSP-templates which include the $g$-, $r$- and $i$-bands. For SN~2012cg, 2012ht and 2013dy
we adopted the standard $R_V = 3.1$ reddening slope (corresponding to the {\tt calibration=2} mode
in SNooPy2), while for SN~2014J, which suffered from strong non-standard reddening, we tested both
the $R_V \sim 1.0$ ({\tt calibration=6}) and $R_V \sim 1.5$ ({\tt calibration=3}) settings. 
These different calibrations are detailed in \citet{fola10}.

All of these codes fit the template LCs to the observed ones via $\chi^2$-minimization, taking into account
photometric errors as inverse weights. The optimized parameters are as follows:
\begin{itemize} 
\item{$T_{max}$: the moment of maximum light in the $B$-band (MJD) }
\item{$A_V^{host}$: the interstellar extinction in the host galaxy in $V$-band (magnitude)}
\item{$\mu_0$: extinction-free distance modulus (magnitude)}
\item{$\Delta$: light curve shape parameter (MLCS2k2)}
\item{$\Delta m_{15}$: light curve shape parameter (SNooPy2)}
\item{$m_B$: peak brightness in $B$-band (SALT2, magnitude)}
\item{$x_0$: light curve normalization parameter (SALT2)}
\item{$x_1$: light curve shape parameter (SALT2)}
\item{$C$: color parameter (SALT2)}
\end{itemize}

Uncertainties of the fitting parameters were calculated
via the standard analysis of the Hessian matrix of the $\chi^2$-hypersurface. 

A particularly important goal of this study was checking the consistency of the distance moduli
obtained from different photometric systems, i.e. to cross-compare the results
from Johnson-Cousins $BVRI$ and Sloan $g'r'i'z'$. SALT2 and SNooPy2 are capable of handling
LCs taken in $g'r'i'$ bands, but not in the $z'$-band. The publicly released version of MLCS2k2
contains only templates in $UBVRI$-bands. In order to make the analysis as complete as possible,
we migrated the MLCS2k2 $UBVRI$ templates to cover the $g'r'i'z'$ filters. Details on this 
step are given in the Appendix.   

Several studies \citep[e.g.][]{hicken09, kelly10, lampeitl10, sulli10} pointed out the 
correlation between SN Ia peak brightnesses and host galaxy stellar masses: 
SNe in more massive ($M_{stellar} \gtrsim 10^{10}$ $M_\odot$) galaxies tend to be slightly
brighter at peak compared to SNe in less massive hosts. 
We discuss the implication of this correlation on the derived distances
in Section~\ref{res}.

Previously the comparison of MLCS2k2 and SALT2 was presented by \citet{kessler09}, who analyzed 
the first-season data from the SDSS-II SN survey. They found moderate disagreement 
(at the $\sim 0.1$ -- $0.2$ mag level) between the distance moduli for their nearby SN sample 
calculated by the two codes. A similar result was obtained by \citet{vinko12} when fitting 
the $BVRI$ data for the extremely well-observed SN~2011fe: the distance moduli given 
by MLCS2k2 and SALT2 differ by $\sim 0.16$ mag. In the rest of the paper we make a similar
comparison for our SN sample by involving SNooPy2 and using data taken in different photometric
systems. 

We would like to emphasize that it is not intended to judge which code is superior over the
others. We use these codes ``as is" without any attempts to fine-tune or retrain their calibrations
to get better match with any particular data, except for bringing them onto the same distance
scale (see below).

\section{Results}\label{res}

This section summarizes the fits obtained from the different methods, and compare the
results with those from previous studies, where applicable. Note that the different
methods use different Hubble-parameters ($H_0$) for their relative distance scales:
MLCS2k2, SNooPy2 and SALT2.4 assume $H_0 = 65$, 72 and 68 km s$^{-1}$ Mpc$^{-1}$, respectively.
In order to bring the reported distance moduli onto the same scale, all of them have been 
corrected to $H_0 = 73$ km s$^{-1}$ Mpc$^{-1}$ \citep{riess16}. The tables below contain these homogenized
distance moduli. The cross-comparison of the distances obtained by these
codes is further discussed in Section~\ref{disc}. 

\subsection{SN~2012cg}

\begin{table}
\caption{Best-fit parameters for SN~2012cg}
\label{tab-12cg-par}
\centering
\begin{tabular}{lcc}
\hline
\hline
Parameter & Value & Error \\
\hline
MLCS2k2 & \multicolumn{2}{c}{$BVRI$}\\
\hline
$R_V$ & 3.1 & fixed \\
$T_{max}$ (MJD) & 56081.5 & 0.30 \\
$A_V^{host}$ (mag) &  0.43 & 0.05 \\
$\Delta$ (mag) & -0.19 & 0.07 \\
$\mu_0$ (mag)  & 30.87 & 0.05 \\
$\chi^2$/d.o.f. & 1.09 & \\
\hline
SNooPy2 & \multicolumn{2}{c}{$BVRI$}\\
\hline
$R_V$ & 3.1 & fixed \\
$T_{max}$ (MJD) & 56082.30 & 0.08 \\
$A_V^{host}$ (mag) & 0.46 & 0.02 \\
$\Delta m_{15}$ (mag) & 0.97 & 0.01\\
$\mu_0$ (mag) & 30.77 & 0.02 \\
$\chi^2$/d.o.f. & 0.93 & \\ 
\hline
SALT2.4 & \multicolumn{2}{c}{$BVRI$}\\
\hline
$T_{max}$ (MJD) & 56082.39 & 0.05 \\
$C$ & 0.08 & 0.02 \\
$x_0$ & 0.28 & 0.01 \\
$x_1$ & 0.45 & 0.04 \\
$m_B$ (mag) & 12.01 & 0.03 \\
$\mu_0$ (mag) & 30.85 & 0.09 \\
$\chi^2$/d.o.f. & 1.44 & \\ 
\hline
\hline
\end{tabular}
\end{table}

Previous photometry of SN~2012cg has been presented and analyzed by \citet{silver12}, 
\citet{muna13}, \cite{ama15} and \citet{marion15b}. 
Table~\ref{tab-12cg-par} lists the optimum parameters found by the 
different LC fitters applied by us. Comparing their values with the ones given by the previous
studies, it is seen that they are generally consistent. There is only a slight tension between
the estimated values of the extinction within the host: the results in Table~\ref{tab-12cg-par} imply
$E(B-V)_{host} = A_V^{host}/R_V = 0.14 \pm 0.02$ mag, while \citet{silver12} and \citet{marion15b}
obtained $E(B-V)_{host} \sim 0.18 \pm 0.05$ mag which is marginally consistent with our results.
Our lower reddening/extinction is closer to $E(B-V)_{host} \sim 0.13$ mag estimated
by \citet{ama15}. For the distance modulus, the parameter that we focus on
in this work, \citet{muna13} obtained $\mu_0 = 30.95$ assuming $E(B-V)=0.18$ mag, which, after
correcting to $E(B-V)=0.14$ mag, corresponds to $\mu_0 = 30.83$ in very good agreement with
our results presented in Table~\ref{tab-12cg-par}.  

The light curves of SN~2012cg are plotted together with the models from the different
LC fitters in Fig.~\ref{fig-12cg}. 
%Fig.~\ref{fig-12cg-mlcs}, \ref{fig-12cg-snpy} and \ref{fig-12cg-salt2}. 

\subsection{SN~2012ht}

\begin{table}
\caption{Best-fit parameters for SN~2012ht}
\label{tab-12ht-par}
\centering
\begin{tabular}{lcccc}
\hline
\hline
Parameter & Value & Error & Value & Error\\
\hline
%\multicolumn{5}{c}{MLCS2k2} \\
%\hline
MLCS2k2 & \multicolumn{2}{c}{$BVRI$} & \multicolumn{2}{c}{$griz$} \\
\hline
$R_V$ & 3.1 & fixed & 3.1 & fixed \\
$T_{max}$ (MJD) & 56295.1 & 0.30 & 56295.10 & 0.30 \\
$A_V^{host}$ (mag) &  0.00 & 0.04 & 0.00 & 0.16\\
$\Delta$ (mag) & 0.24 & 0.04 & 0.28 & 0.12 \\
$\mu_0$ (mag)  & 32.16 & 0.04 & 32.11 & 0.14 \\
$\chi^2$/d.o.f. & 0.91 & & 1.73 & \\ 
\hline
%\multicolumn{5}{c}{SNooPy2}\\
SNooPy2 & \multicolumn{2}{c}{$BVRI$} & \multicolumn{2}{c}{$gri$} \\
\hline
$R_V$ & 3.1 & fixed & 3.1 & fixed\\
$T_{max}$ (MJD) & 56295.23 & 0.08 & 56295.00 & 0.21 \\
$A_V^{host}$ (mag) & 0.00 & 0.01 & 0.00 & 0.01 \\
$\Delta m_{15}$ (mag) & 1.30 & 0.01 & 1.17 & 0.01\\
$\mu_0$ (mag) & 32.26 & 0.02 & 32.44 & 0.02\\ 
$\chi^2$/d.o.f. & 0.58 & & 1.16 & \\ 
\hline
%\multicolumn{5}{c}{SALT2.4}\\
SALT2.4 & \multicolumn{2}{c}{$BVRI$} & \multicolumn{2}{c}{$gri$} \\
\hline
$T_{max}$ (MJD) & 56295.47 & 0.07 & 56295.50 & 0.30 \\
$C$ & -0.08 & 0.03 & -0.14 & 0.05 \\
$x_0$ & 0.111 & 0.003 & 0.116 & 0.007 \\
$x_1$ & -1.25 & 0.05 & -0.81 & 0.26 \\
$m_B$ (mag) & 13.02 & 0.03 & 12.98 & 0.07 \\
$\mu_0$ (mag) & 32.12 & 0.09 & 32.32 & 0.17 \\
$\chi^2$/d.o.f. & 1.47 & & 2.02 & \\  
\hline
\hline
\end{tabular}
\end{table}

\begin{figure*}
\centering
\resizebox{8cm}{!}{\includegraphics{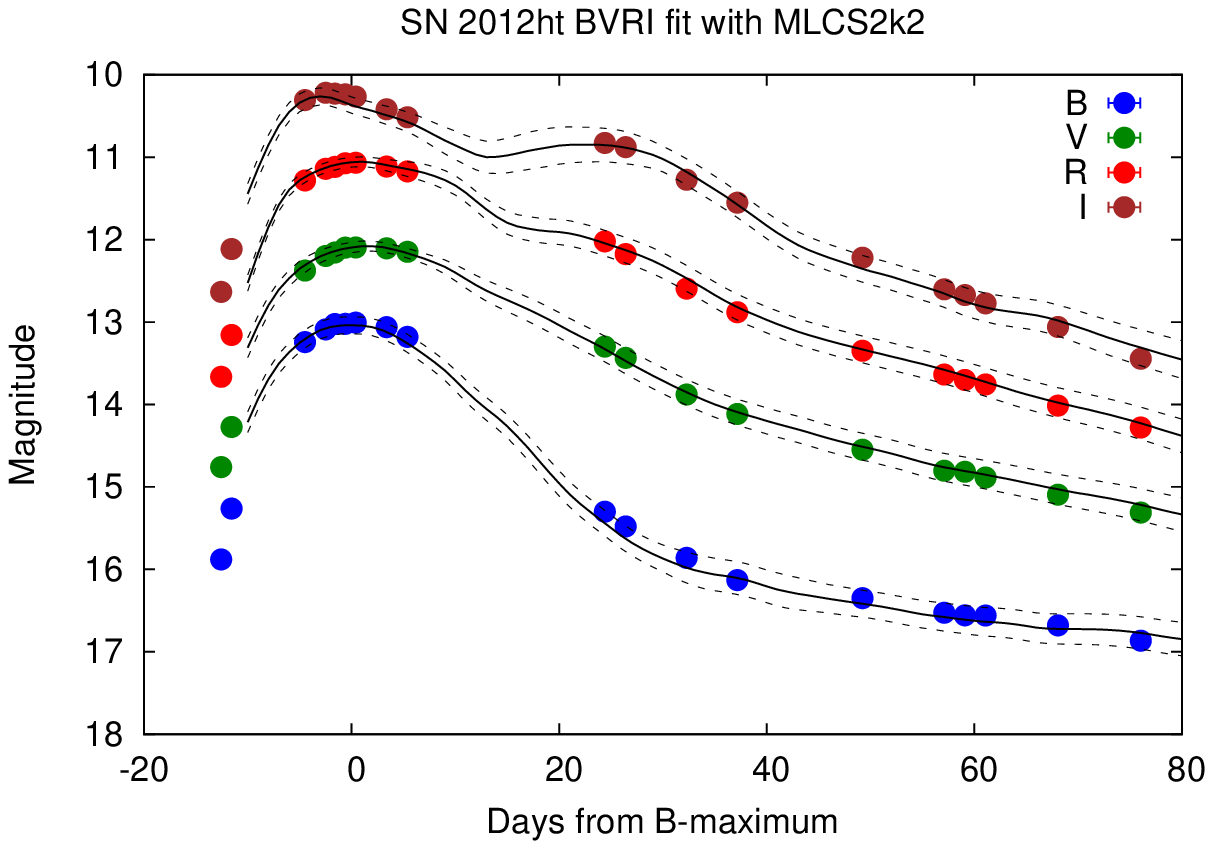}}
\resizebox{8cm}{!}{\includegraphics{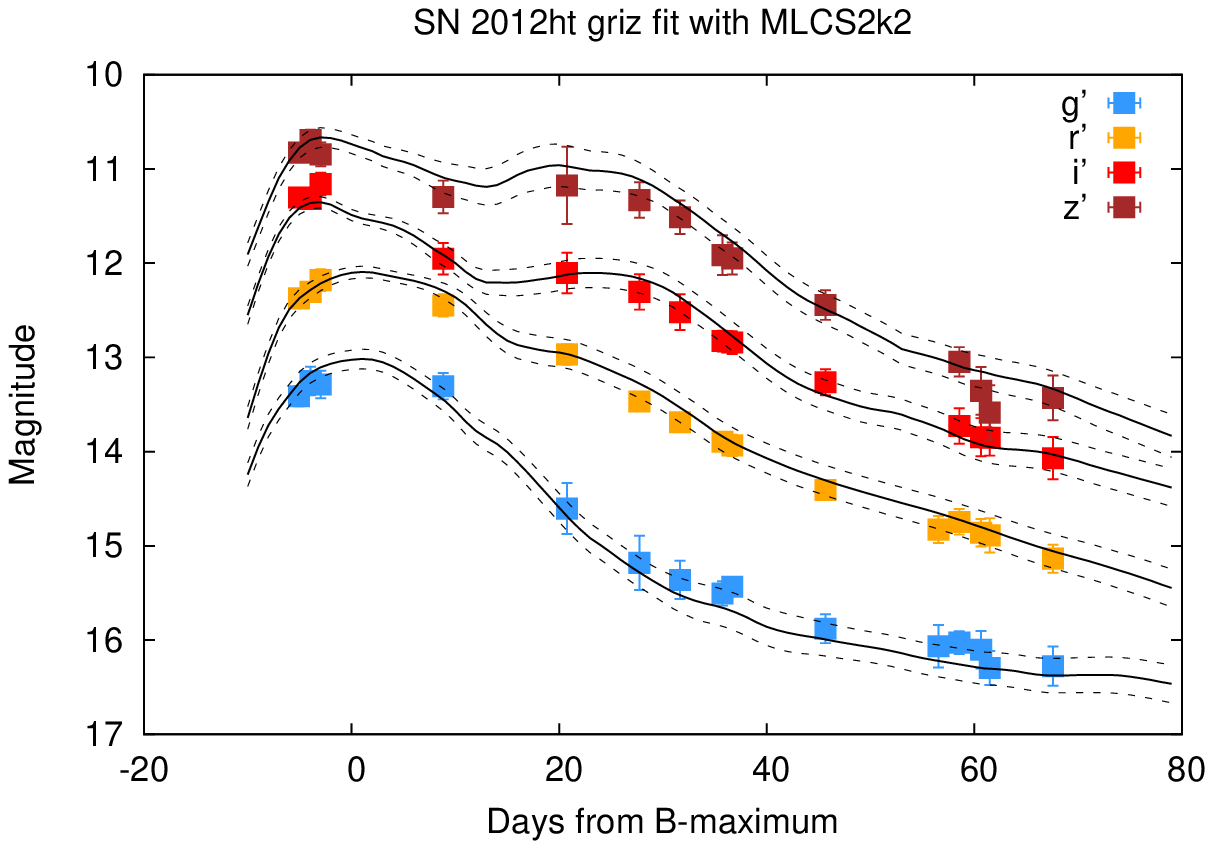}}\\
\resizebox{8cm}{!}{\includegraphics{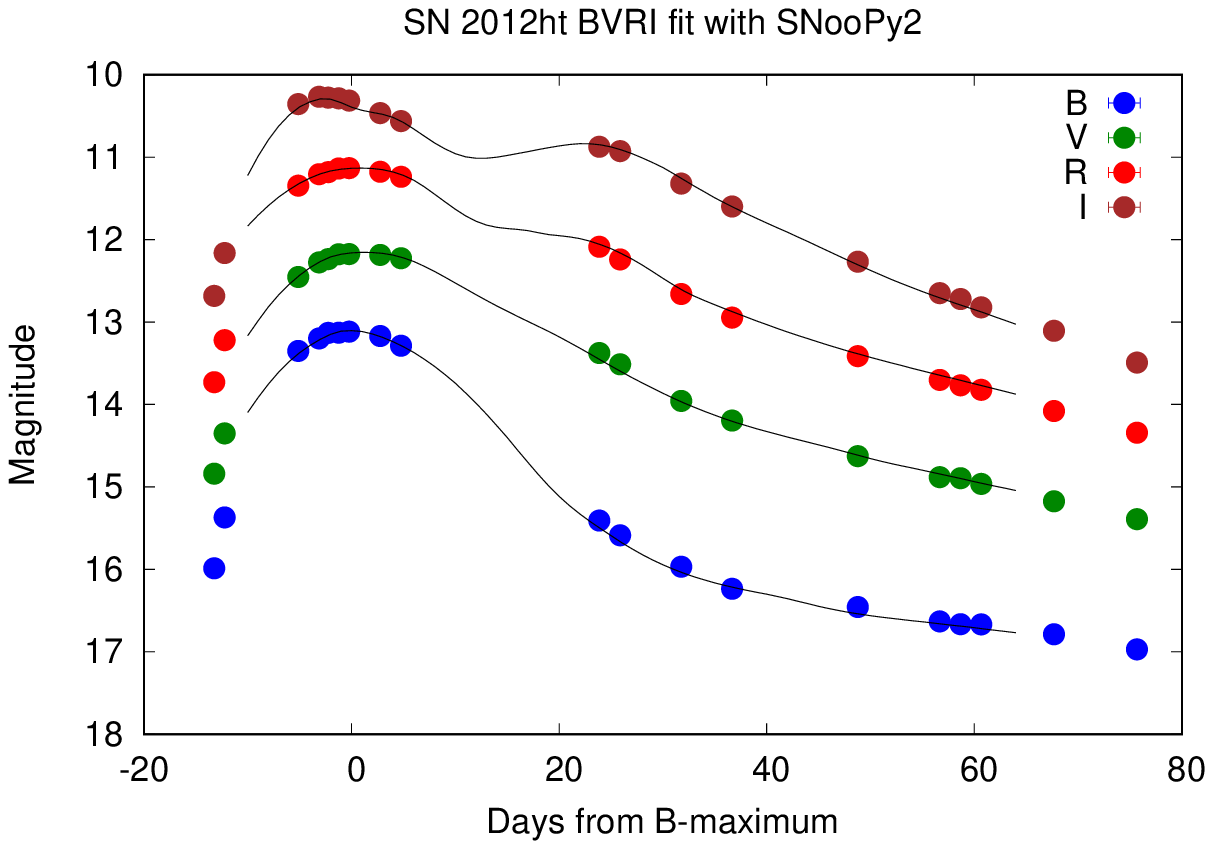}}
\resizebox{8cm}{!}{\includegraphics{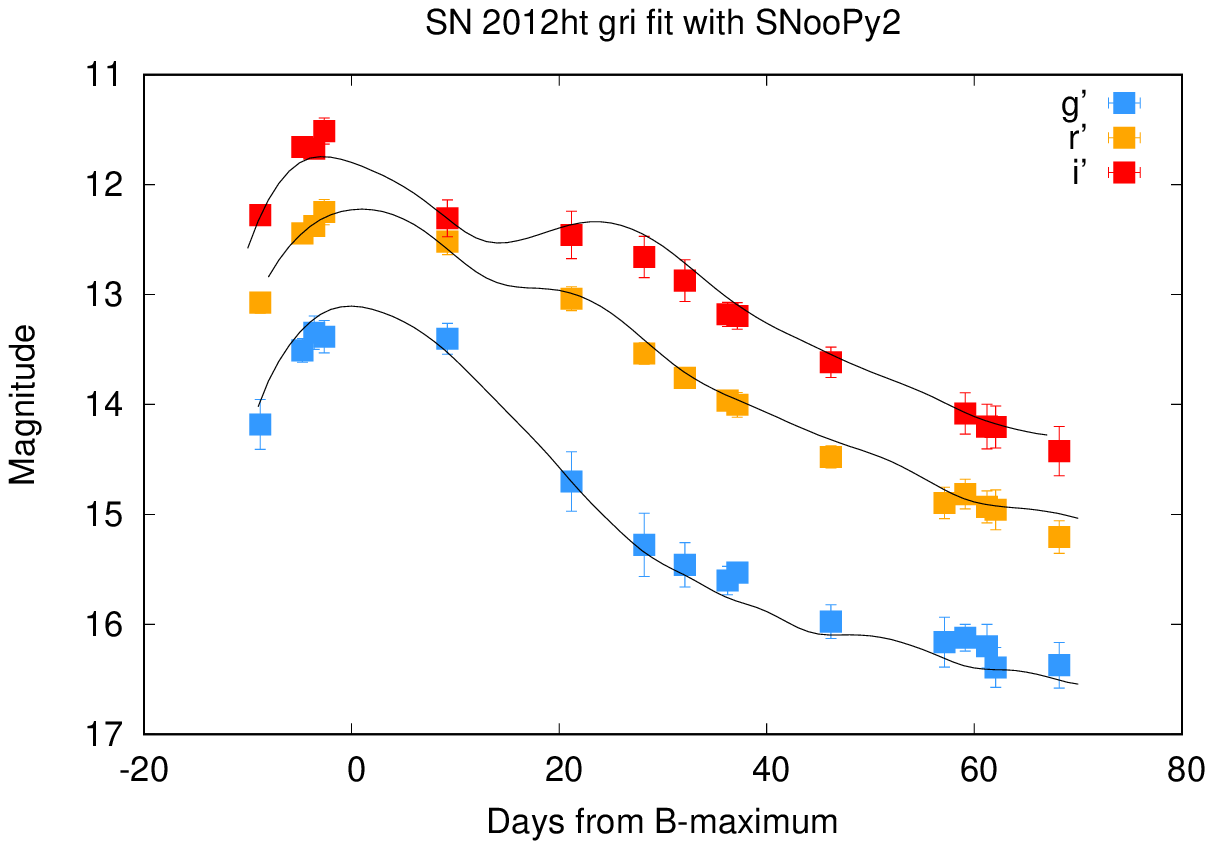}}\\
\resizebox{8cm}{!}{\includegraphics{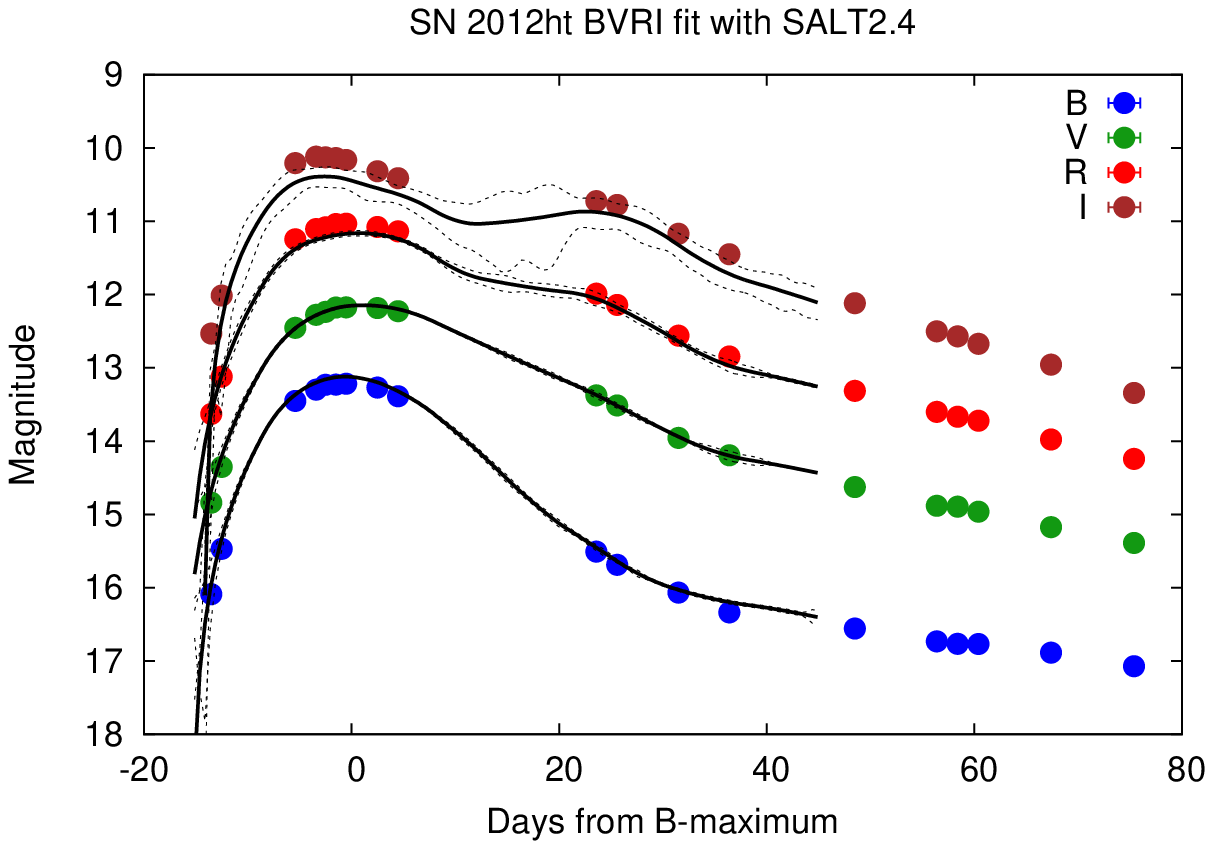}}
\resizebox{8cm}{!}{\includegraphics{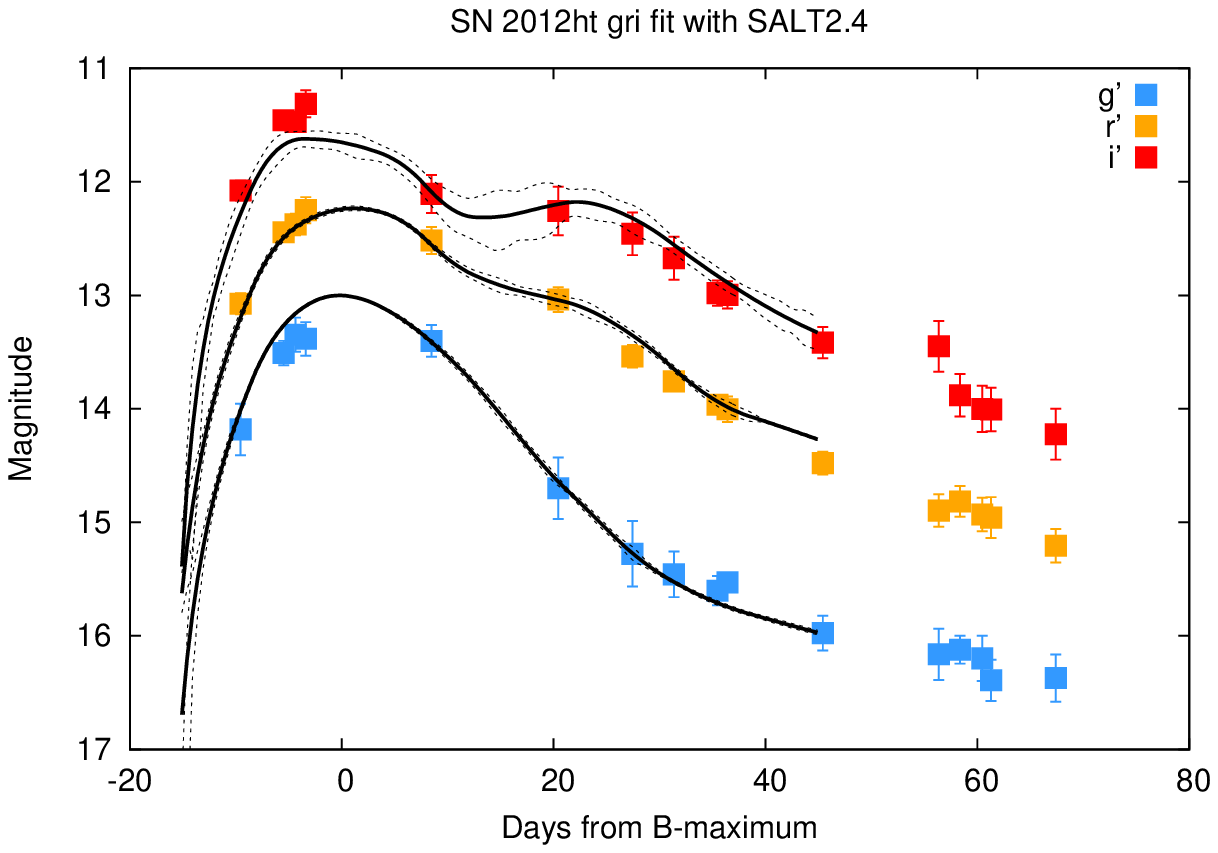}}
\caption{The fitting of the light curves of SN~2012ht, after correcting for Milky Way extinction. 
Top row: MLCS2k2; middle row: SNooPy2; bottom row: SALT2.4;
left column: $BVRI$ data; right column: $g'r'i'z'$ data. }
\label{fig-12ht}
\end{figure*}

A photometric study of SN~2012ht has been presented by \citet{yama14}. From their $BVRI$ 
photometry they have estimated the following parameters: $T_{max}(B) = 56295.6 \pm 0.6$, 
$\Delta m_{15}(B) = 1.39$ mag and $E(B-V)_{host} \sim 0$ mag. 
As seen from Table~\ref{tab-12ht-par}, these are in good agreement with our
results. The light curves together with the best-fit models can be found in 
Fig.~\ref{fig-12ht}. It is seen that the $BVRI$ data that have lower measurement
errors could be fit better: their reduced $\chi^2$ values (Table~\ref{tab-12ht-par}) 
are lower than those of the $g'r'i'z'$ data.

In order to test the consistency of the photometric calibration of 
our $BVRI$ and $g'r'i'$ data, simultaneous fits to the combined LCs
were also computed with SNooPy2 and SALT2.4 (MLCS2k2 was trained only on $BVRI$ data,
so that code was not applied in this test). As expected, these fits 
produced slightly higher $\chi^2$ values than the fits to the $BVRI$ LCs
alone, but their best-fit parameters were consistent with
the ones listed in Table~\ref{tab-12ht-par}. In particular, the distance
modulus from the combined fits turned out to be $\mu_0 = 32.20 \pm 0.01$ mag
($\chi^2$/d.o.f=0.83) from SNooPy2, while from SALT2.4 it is $32.13 \pm 0.08$ mag 
($\chi^2$/d.o.f.=2.81). These parameters are closer to those obtained from fitting
the $BVRI$ LCs than those from fitting the $g'r'i'z'$ data, probably because of
the lower measurement uncertainties of the former.  

\subsection{SN~2013dy}

\begin{table}
\caption{Best-fit parameters for SN~2013dy}
\label{tab-13dy-par}
\centering
\begin{tabular}{lcccc}
\hline
\hline
Parameter & Value & Error & Value & Error\\
\hline
%\multicolumn{5}{c}{MLCS2k2} \\
%\hline
MLCS2k2 & \multicolumn{2}{c}{$BVRI$} & \multicolumn{2}{c}{$griz$} \\
\hline
$R_V$ & 3.1 & fixed & 3.1 & fixed \\
$T_{max}$ (MJD) & 56500.20 & 0.30 & 56500.20 & 0.30 \\
$A_V^{host}$ (mag) & 0.48 & 0.06 & 0.28 & 0.16 \\
$\Delta$ (mag) & -0.23 & 0.06 & -0.30 & 0.12 \\
$\mu_0$ (mag)  & 31.51 & 0.06 & 31.65 & 0.13 \\
$\chi^2$/d.o.f. & 1.34 & & 1.91 & \\ 
\hline
%\multicolumn{5}{c}{SNooPy2}\\
SNooPy2 & \multicolumn{2}{c}{$BVRI$} & \multicolumn{2}{c}{$gri$} \\
\hline
$R_V$ & 3.1 & fixed & 3.1 & fixed\\
$T_{max}$ (MJD) & 56501.30 & 0.08 & 56501.03 & 0.15 \\
$A_V^{host}$ (mag) & 0.40 & 0.02 & 0.61 & 0.05 \\
$\Delta m_{15}$ (mag) & 0.96 & 0.01 & 0.80 & 0.01\\
$\mu_0$ (mag) & 31.52 & 0.03 & 31.44 & 0.04\\ 
$\chi^2$/d.o.f. & 0.99 & & 3.45 & \\ 
\hline
%\multicolumn{5}{c}{SALT2.4}\\
SALT2.4 & \multicolumn{2}{c}{$BVRI$} & \multicolumn{2}{c}{$gri$} \\
\hline
$T_{max}$ (MJD) & 56501.44 & 0.06 & 56501.99 & 0.14 \\
$C$ & 0.089 & 0.025 & 0.03 & 0.03 \\
$x_0$ & 0.154 & 0.004 & 0.169 & 0.006 \\
$x_1$ & 0.695 & 0.044 & 1.51 & 0.12 \\
$m_B$ (mag) & 12.67 & 0.03 & 12.57 & 0.04 \\
$\mu_0$ (mag) & 31.52 & 0.09 & 31.72 & 0.11 \\ 
$\chi^2$/d.o.f. & 0.76 & & 2.70 & \\ 
\hline
\hline
\end{tabular}
\end{table}

\begin{figure*}
\centering
\resizebox{8cm}{!}{\includegraphics{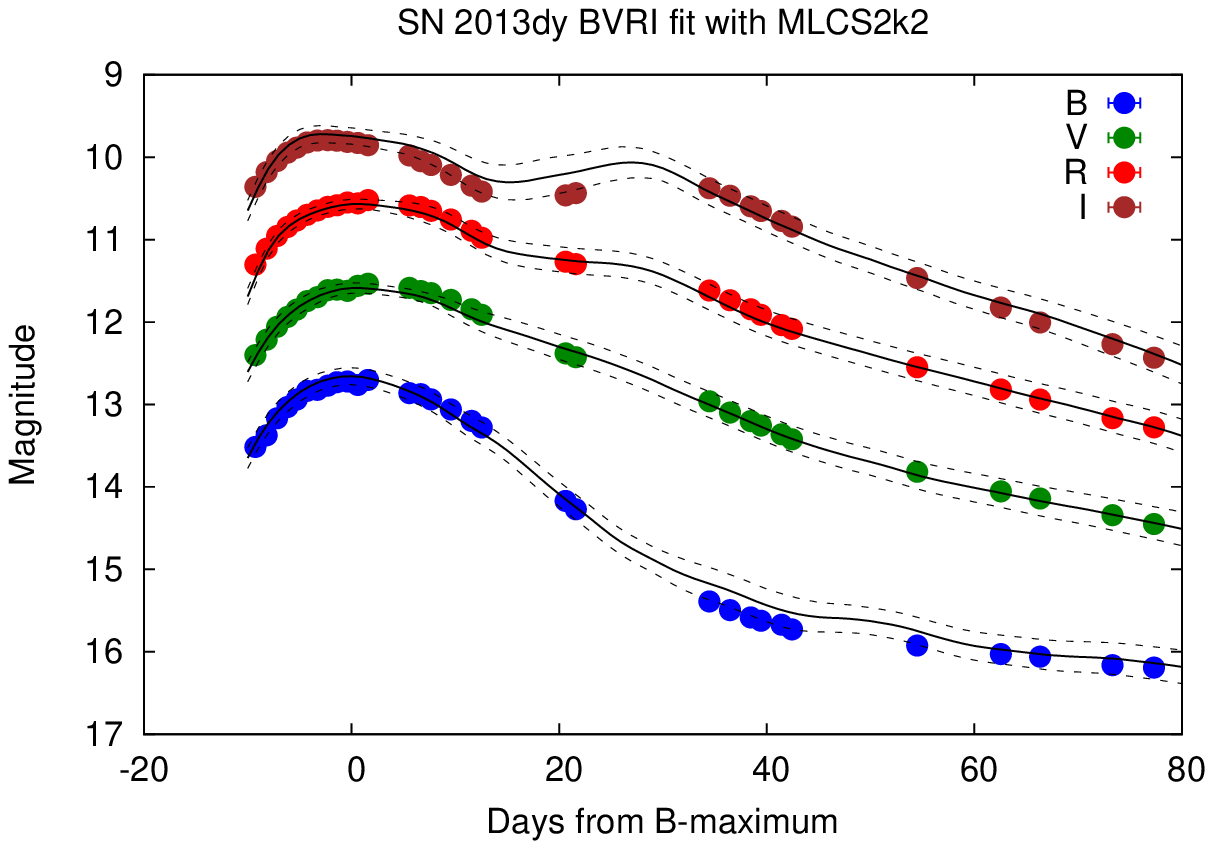}}
\resizebox{8cm}{!}{\includegraphics{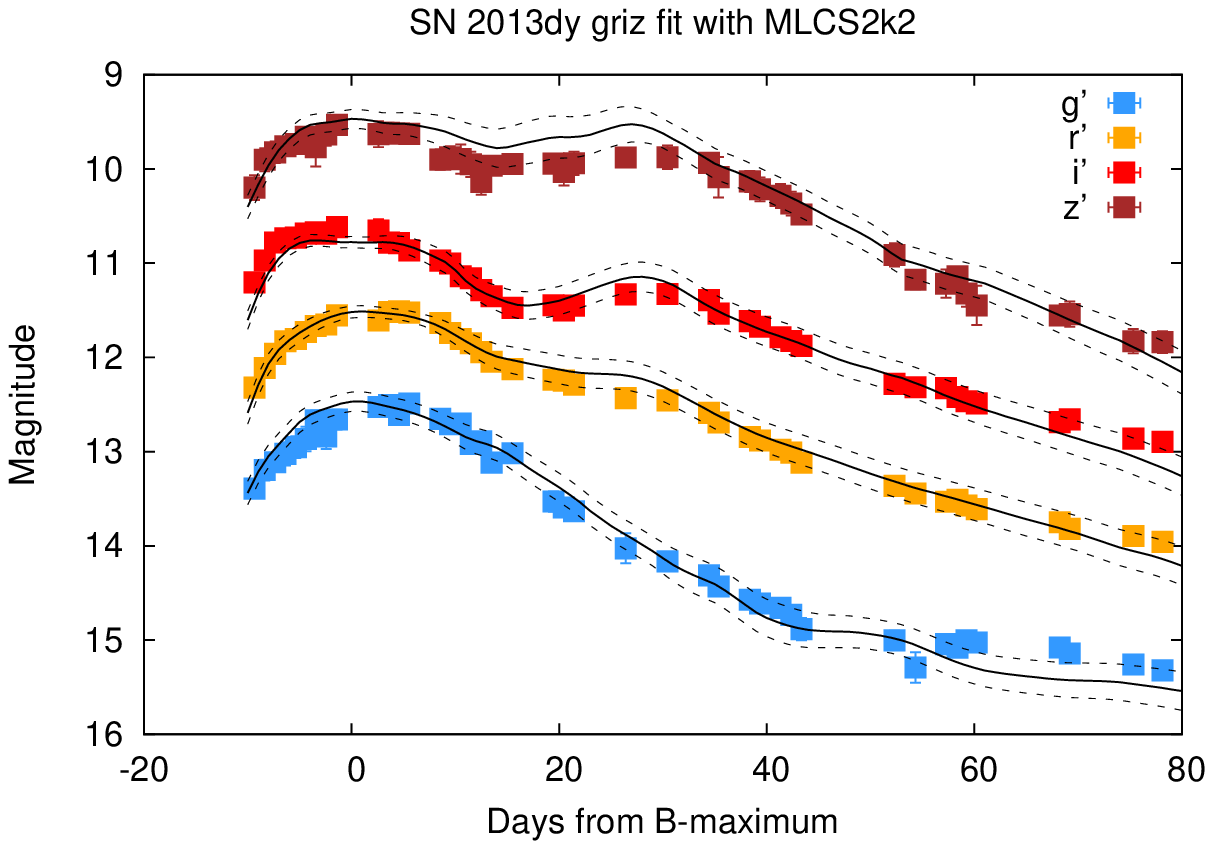}}\\
\resizebox{8cm}{!}{\includegraphics{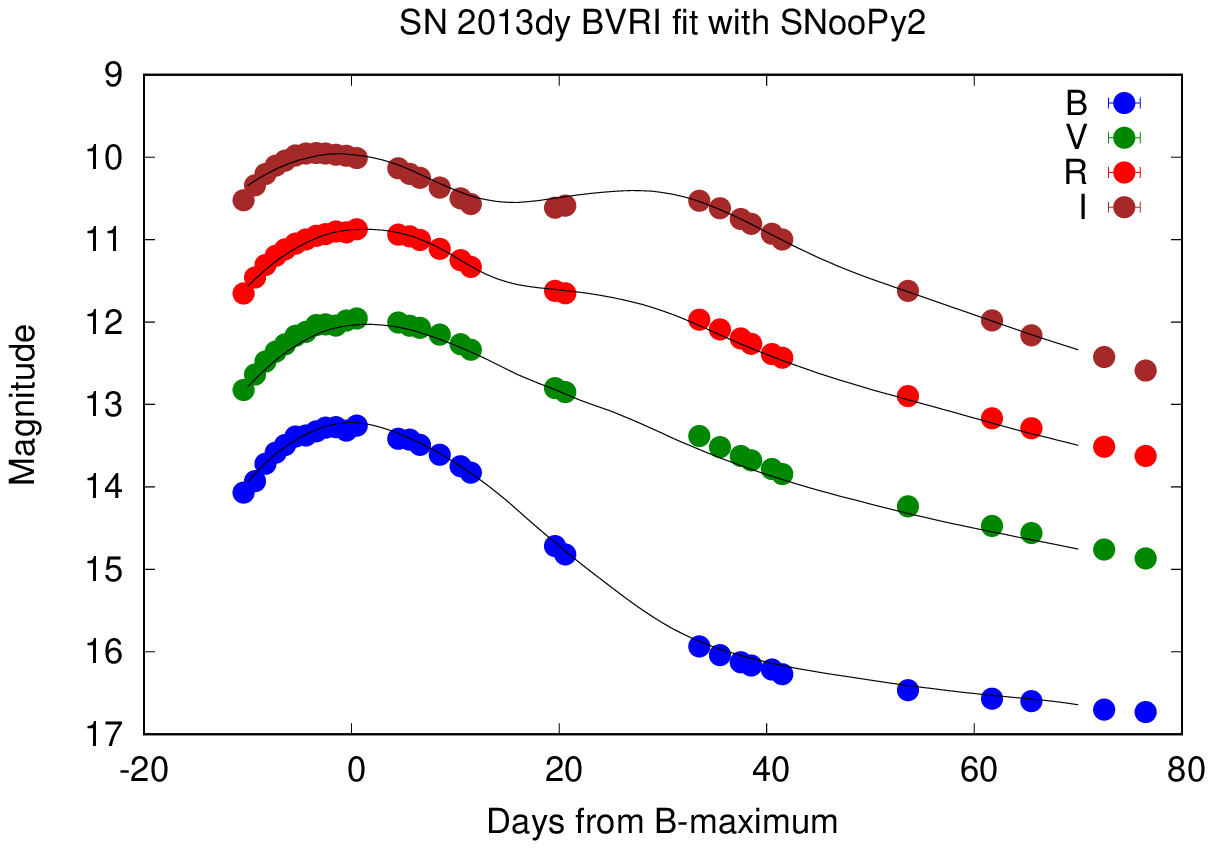}}
\resizebox{8cm}{!}{\includegraphics{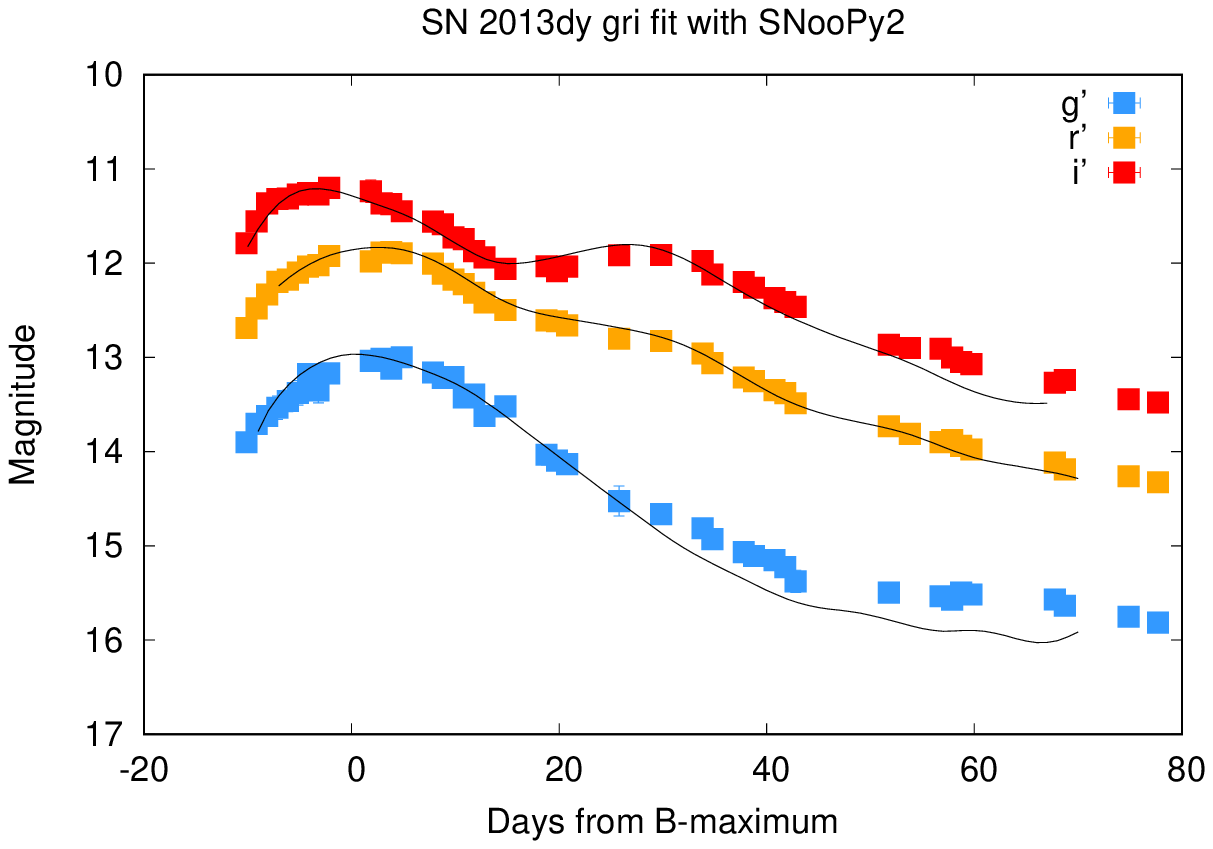}}\\
\resizebox{8cm}{!}{\includegraphics{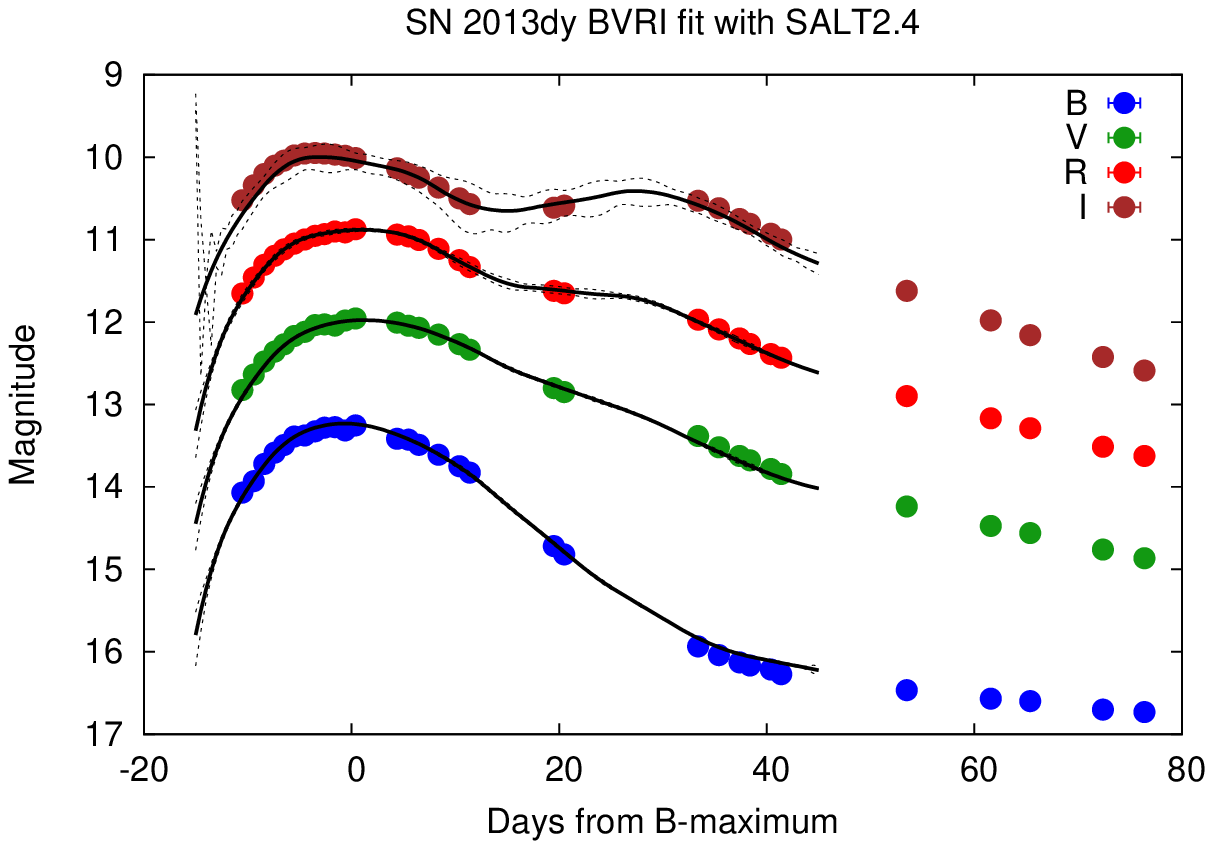}}
\resizebox{8cm}{!}{\includegraphics{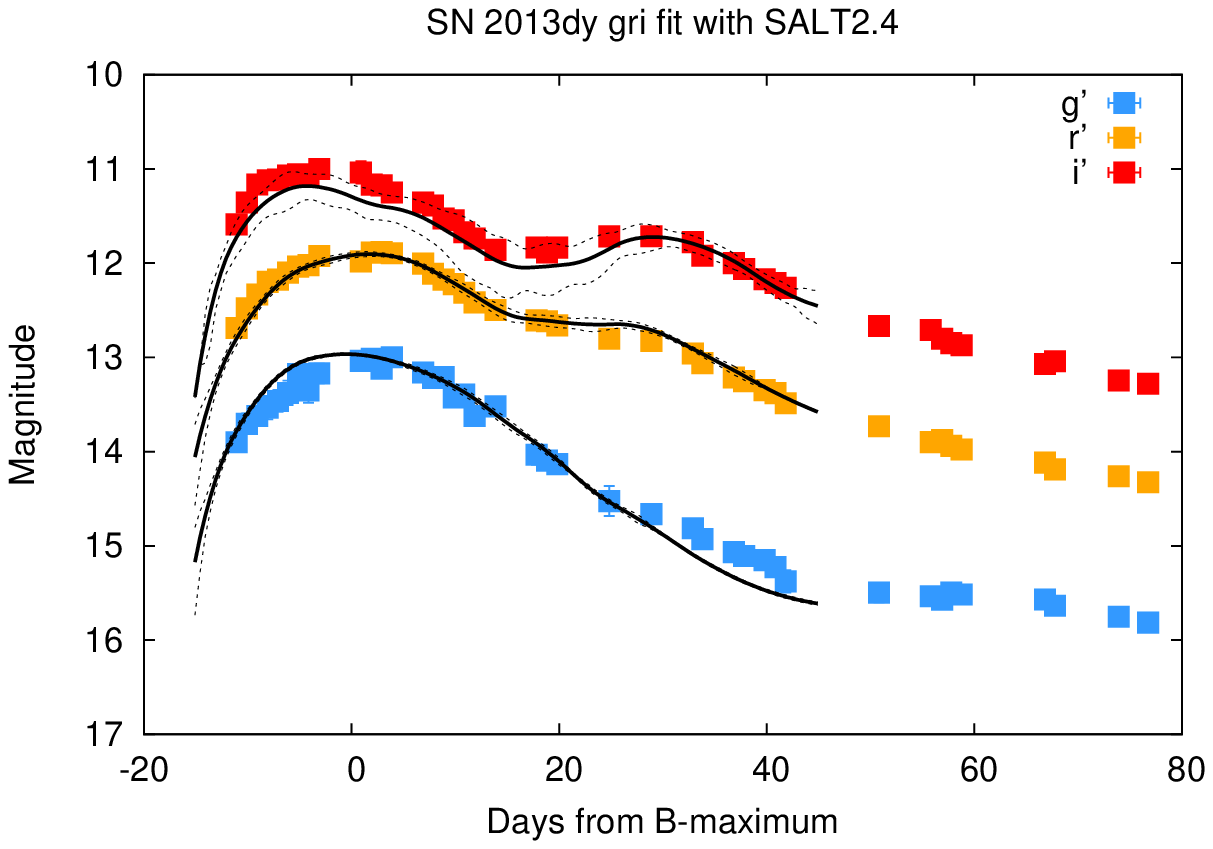}}
\caption{The same as Fig.\ref{fig-12ht} but for SN~2013dy.}
\label{fig-13dy}
\end{figure*}

Light curves of SN~2013dy have been published recently by \citet{pan15} (P15) and \citet{zhai16} (Z16) in the $BVRIriZYJH$ and
$UBVRI$ bands, respectively. \citet{zhai16} also presented photometry obtained by {\it Swift/UVOT}.   

\citet{pan15} applied the SNooPy2 code to fit their full $BVRIriZYJH$ dataset simultaneously, 
and obtained $T_{max}(B) = 56501.1$, $\Delta m_{15} = 0.886 \pm 0.006$, 
$E(B-V)_{host} = 0.206 \pm 0.005$ mag and $\mu_0 = 31.49 \pm 0.01$ mag. 
Comparing these values with those in Table~\ref{tab-13dy-par} it is apparent
that the results of \citet{pan15} are close to the ones obtained in the
present study, although the differences somewhat exceed the formal errors
given by SNooPy2. Comparing the best-fit values of the common parameters
obtained from different methods, e.g. $A_V^{host}$ or $\mu_0$, it is seen that
they also deviate much more than the uncertainties given by the codes. 
Thus, it is suspected that the formal parameter errors, especially
those reported by SNooPy2, are underestimated, and the true uncertainties
should be higher. Keeping this in mind,
the solutions presented in Table~\ref{tab-13dy-par} are entirely consistent
with the LC fit given by \citet{pan15}. The fit of the model LCs to the
data can be seen in Fig.~\ref{fig-13dy}.

\begin{table}
\caption{Cross-comparison of the parameters from fitting independent data on SN~2013dy. 
Uncertainties are in parentheses.}
\label{tab-13dy-xcheck}
\centering
\begin{tabular}{llll}
\hline
\hline
Parameter & this work & P15 & Z16 \\
\hline
\multicolumn{4}{l}{MLCS2k2:}\\
$T_{max}$ & 56500.2 (0.3) & 56500.5 (0.3) & 56501.4 (0.3) \\
$A_V^{host}$ & 0.48 (0.06) & 0.42 (0.07) & 0.45 (0.06) \\
$\Delta$ & $-0.23$ (0.06) & $-0.20$ (0.05) & $-0.22$ (0.05) \\
$\mu_0$ & 31.51 (0.06) & 31.53 (0.06) & 31.53 (0.07) \\
\hline
\multicolumn{4}{l}{SNooPy2:}\\
$T_{max}$ & 56501.30 (0.08) & 56501.48 (0.11) & 56501.29 (0.11) \\
$A_V^{host}$ & 0.40 (0.02) & 0.31 (0.02) & 0.43 (0.02) \\
$\Delta m_{15}$ & 0.96 (0.01) & 0.97 (0.02) & 0.86 (0.02) \\
$\mu_0$ & 31.52 (0.03) & 31.56 (0.01) & 31.52 (0.02) \\
\hline
\multicolumn{4}{l}{SALT2.4:}\\
$T_{max}$ & 56501.44 (0.06) & 56501.47 (0.04) & 56502.09 (0.14) \\
$C$ & 0.089 (0.025) & 0.081 (0.016) & 0.149 (0.024) \\
$x_0$ & 0.154 (0.004) & 0.152 (0.003) & 0.143 (0.004) \\
$x_1$ & 0.695 (0.044) & 0.814 (0.044) & 1.002 (0.073) \\
$m_B$ & 12.670 (0.028) & 12.682 (0.025) & 12.743 (0.028) \\
$\mu_0$ & 31.52 (0.08) & 31.573 (0.069) & 31.450 (0.090) \\
\hline
\hline
\end{tabular}
\end{table}

We re-analyzed the $BVRI$ light curves of SN~2013dy from both \citet{pan15} and \citet{zhai16} with all three
methods applied in this paper in order to cross-compare the results from fitting measurements
taken independently on the same SN. We assumed $R_V = 3.1$ for all fits, as earlier. 
The results are shown in Table~\ref{tab-13dy-xcheck}. It is seen that the consistency between the 
distance moduli from the three methods is excellent for all data. Overall, the distance moduli from the Konkoly 
data differ by less than $\sim 1 \sigma$ ($\lesssim 0.1$ mag) from both the P15 and Z16 results. 
Also, it is seen from Table~\ref{tab-13dy-xcheck} that similar amount of dispersion in the distance
moduli derived by the different codes from the same dataset appears for the P15 and Z16 LCs as well as
for our data, which suggests that this dispersion is probably not simply due to photometric calibration 
issues, at least for the $BVRI$ data.    

In addition, we also modeled the combined $BVRI$ + $g'r'i$ LCs as in the case for SN~2012ht. 
The resulting distance moduli are $\mu_0 = 31.45 \pm 0.01$ mag ($\chi^2$/d.o.f.=2.92) from
SNooPy2 and $\mu_0 = 31.54 \pm 0.07$ mag ($\chi^2$/d.o.f.=2.11) from SALT2.4. Again, these
results are in good agreement with those listed in Table~\ref{tab-13dy-par}, even though the
$\chi^2$ values of the combined fits are somewhat higher but still acceptable (note again 
that the uncertainty of $\mu_0$ reported by SNooPy2 is underestimated).

\subsection{SN~2014J}

\begin{figure*}
\centering
\resizebox{8cm}{!}{\includegraphics{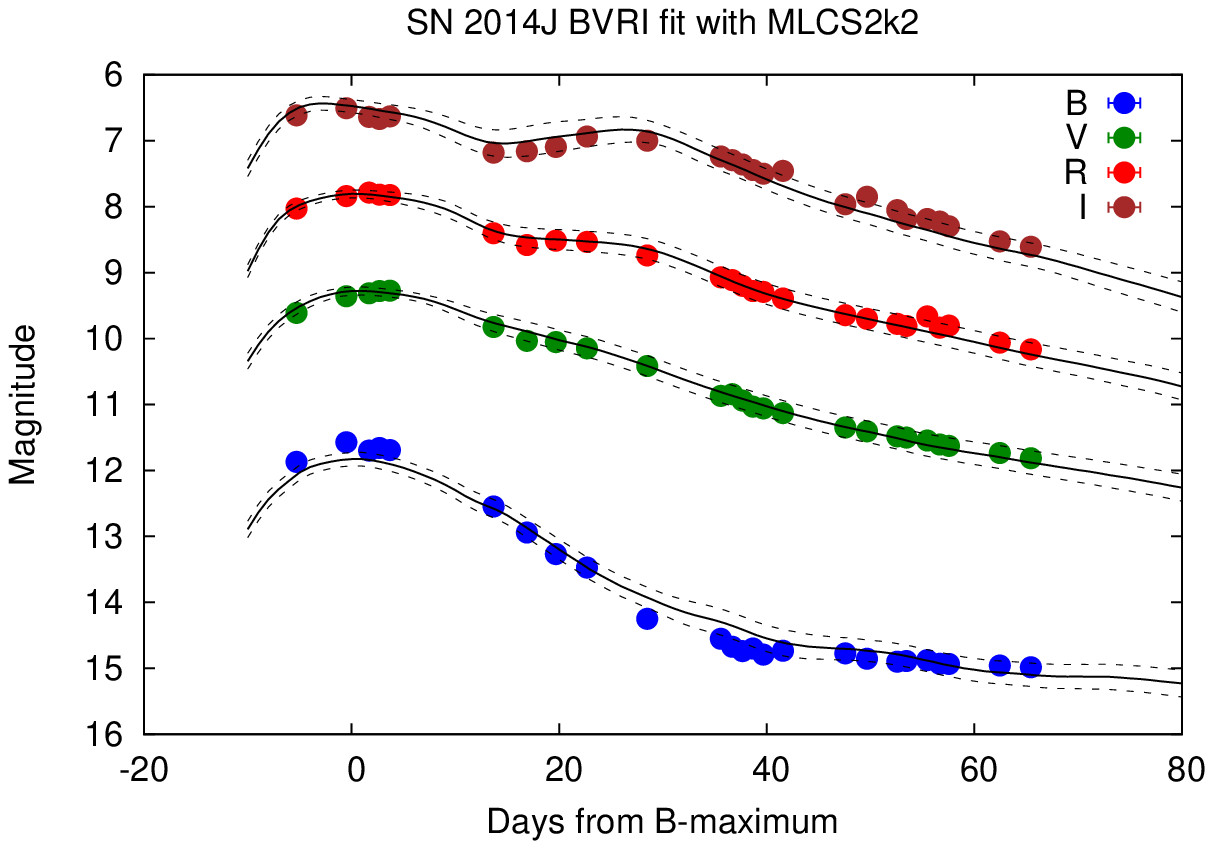}}
\resizebox{8cm}{!}{\includegraphics{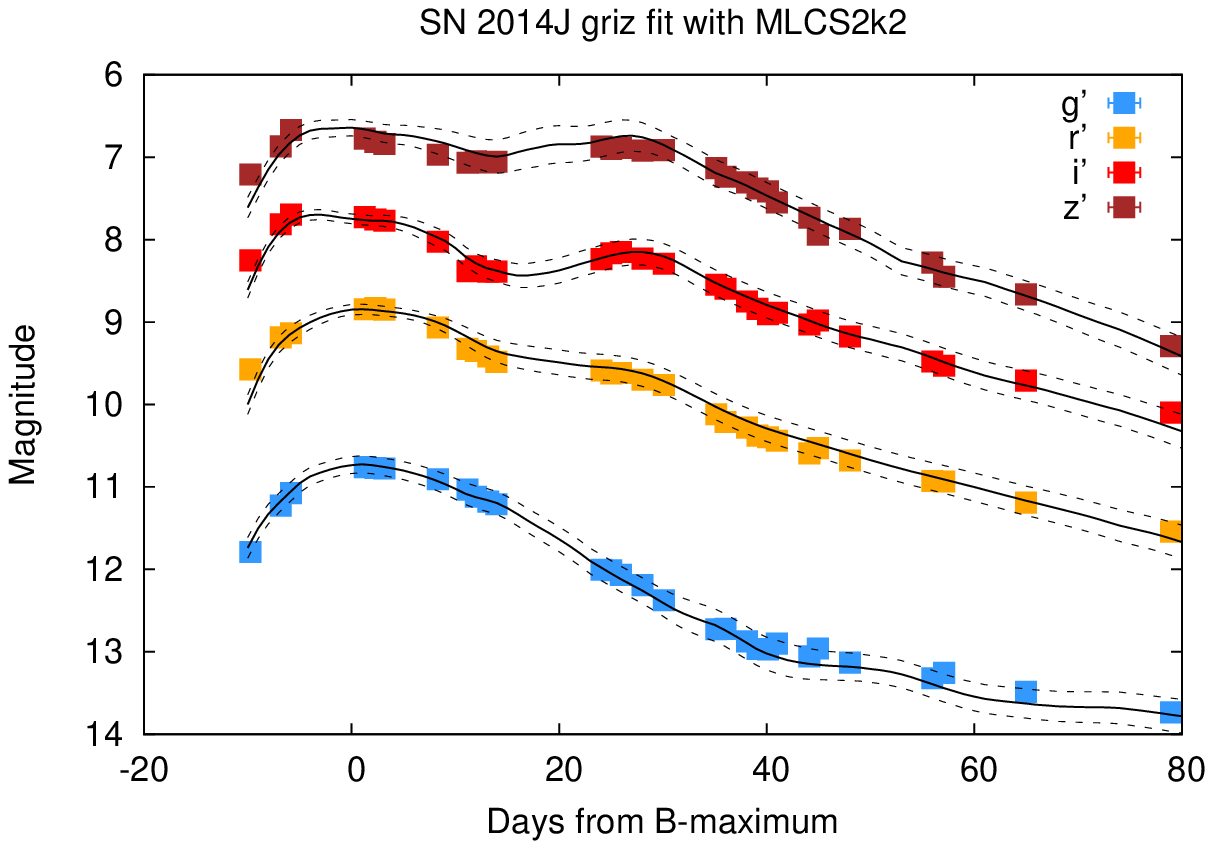}}\\
\resizebox{8cm}{!}{\includegraphics{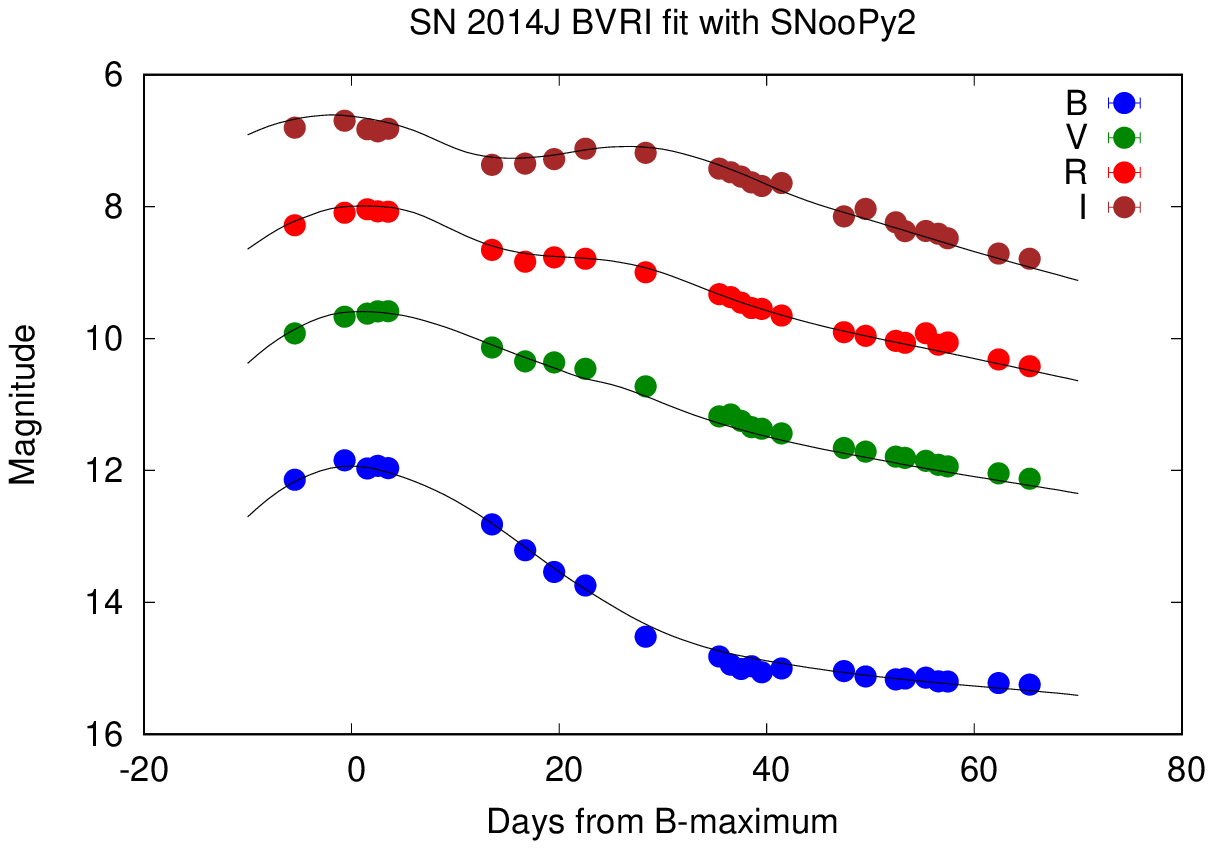}}
\resizebox{8cm}{!}{\includegraphics{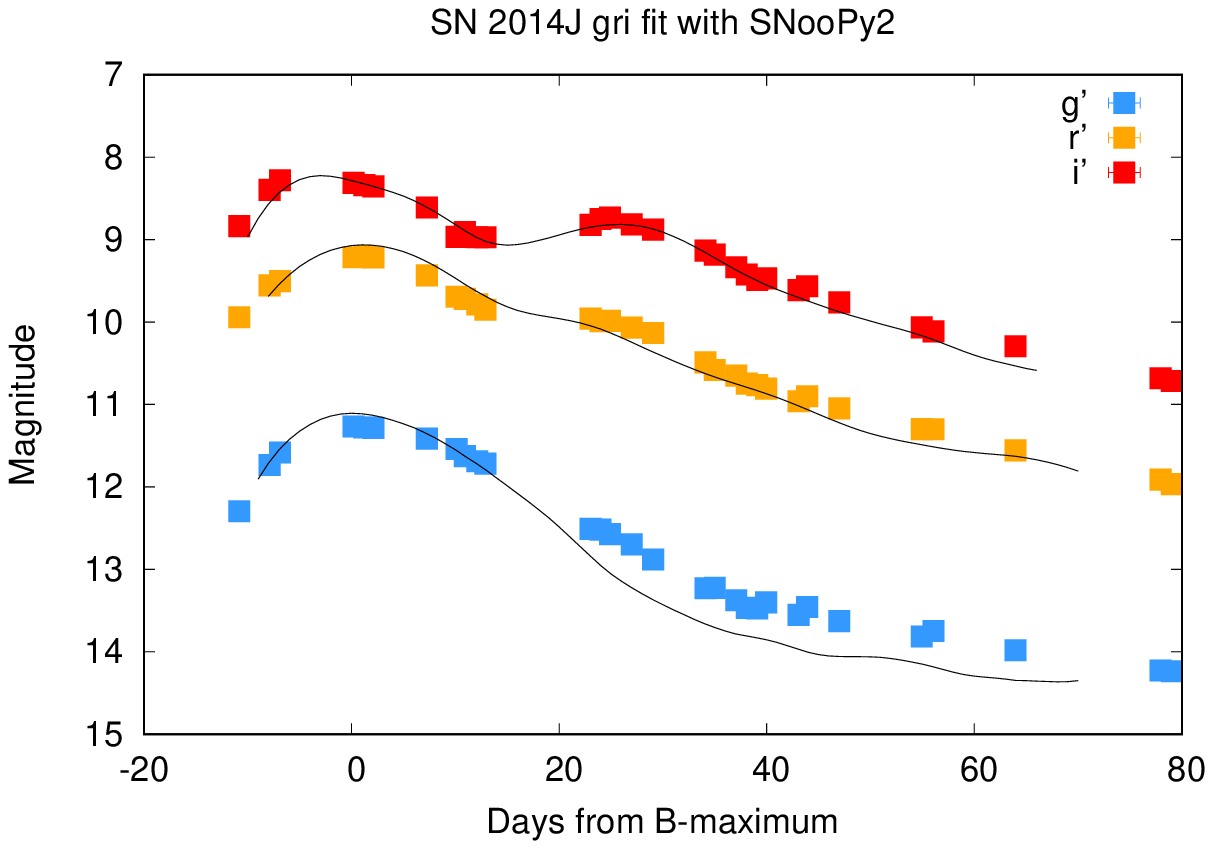}}\\
\resizebox{8cm}{!}{\includegraphics{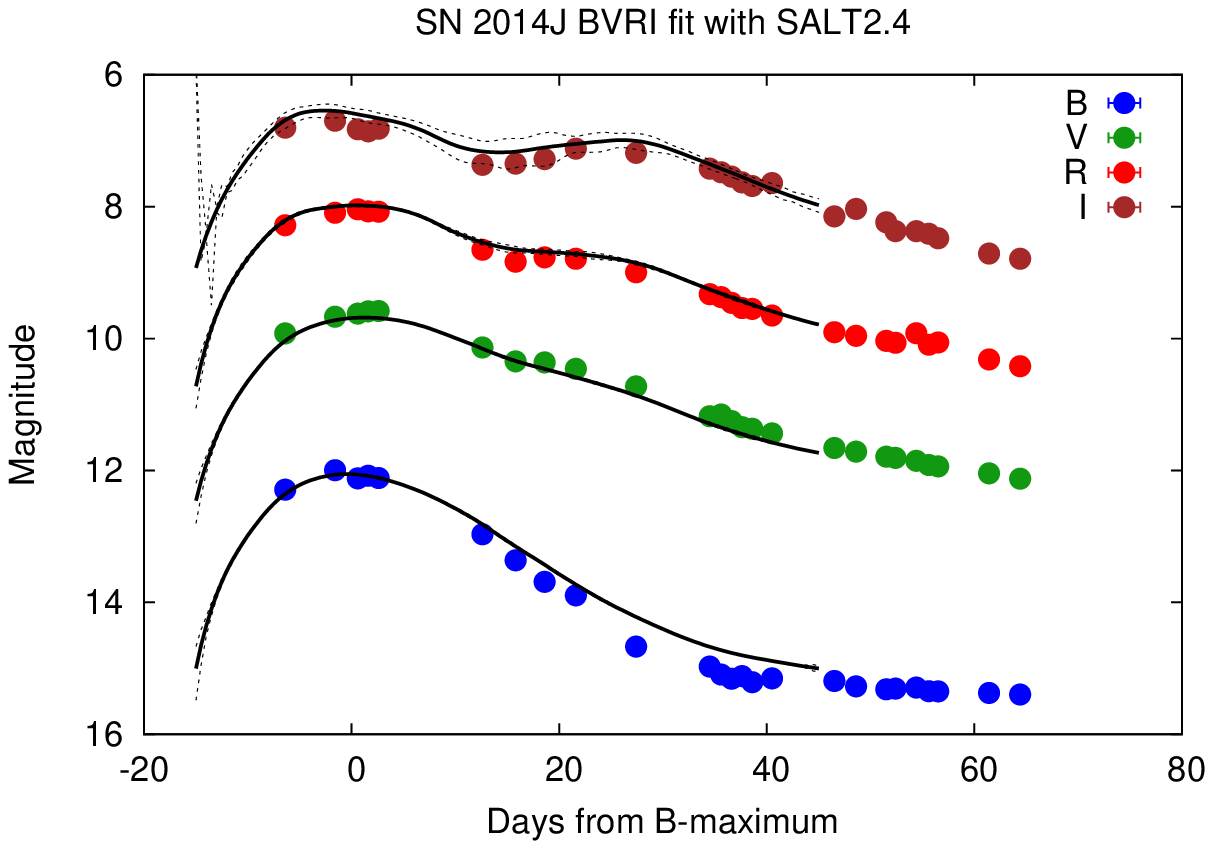}}
\resizebox{8cm}{!}{\includegraphics{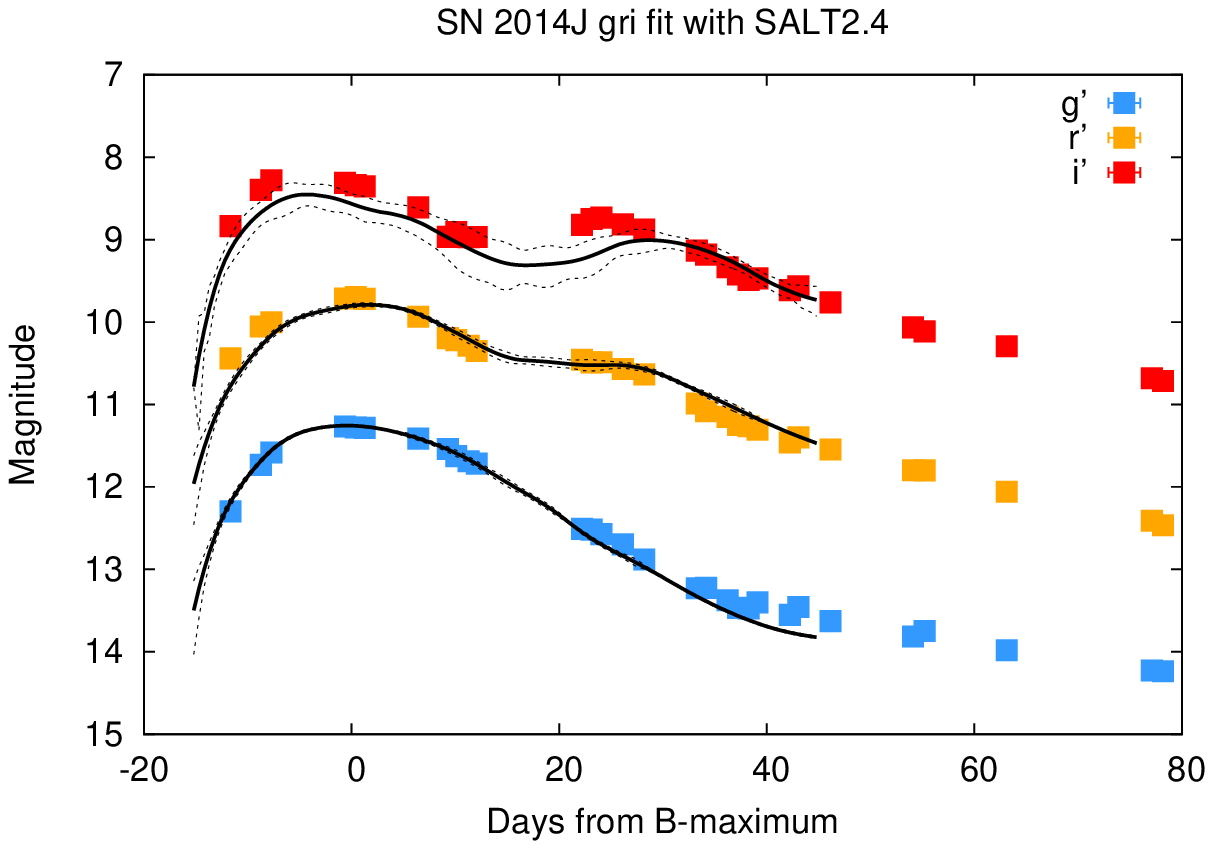}}
\caption{The same as Fig.\ref{fig-12ht} but for SN~2014J.}
\label{fig-14J}
\end{figure*}

\begin{table}
\caption{Best-fit parameters for SN~2014J}
\label{tab-14J-par}
\centering
\begin{tabular}{lcccc}
\hline
\hline
Parameter & Value & Error & Value & Error\\
\hline
%\multicolumn{5}{c}{MLCS2k2} \\
%\hline
MLCS2k2 & \multicolumn{2}{c}{$BVRI$} & \multicolumn{2}{c}{$griz$} \\
\hline
$R_V$ & 1.4 & fixed & 1.4 & fixed \\
$T_{max}$ (MJD) & 56689.8 & 0.50 & 56689.8 & 0.50 \\
$A_V^{host}$ (mag) & 1.84 & 0.09 & 1.49 & 0.16 \\
$\Delta$ (mag) & -0.13 & 0.08 & -0.24 & 0.12 \\
$\mu_0$ (mag)  & 27.72 & 0.09 & 27.99 & 0.13 \\
$\chi^2$/d.o.f. & 3.62 & & 2.37 & \\ 
\hline
$R_V$ & 1.0 & fixed & 1.0 & fixed \\
$T_{max}$ (MJD) & 56689.8 & 0.50 & 56689.8 & 0.50 \\
$A_V^{host}$ (mag) & 1.41 & 0.08 & 1.16 & 0.16 \\
$\Delta$ (mag) & -0.16 & 0.07 & -0.25 & 0.12 \\
$\mu_0$ (mag)  & 28.14 & 0.08 & 28.21 & 0.13 \\
$\chi^2$/d.o.f. & 3.15 & & 2.32 & \\ 
\hline
%\multicolumn{5}{c}{SNooPy2}\\
SNooPy2 & \multicolumn{2}{c}{$BVRI$} & \multicolumn{2}{c}{$gri$} \\
\hline
$R_V$ & 1.5 & fixed & 1.5 & fixed \\
$T_{max}$ (MJD) & 56689.99 & 0.42 & 56690.86 & 0.24 \\
$A_V^{host}$ (mag) & 1.89 & 0.05 & 1.40 & 0.03 \\
$\Delta m_{15}$ (mag) & 1.04 & 0.04 & 1.05 & 0.03\\
$\mu_0$ (mag) & 27.50 & 0.02 & 27.69 & 0.05\\
$\chi^2$/d.o.f. & 4.21 & & 16.73 & \\  
\hline
$R_V$ & 1.0 & fixed & 1.0 & fixed \\
$T_{max}$ (MJD) & 56689.99 & 0.34 & 56690.86 & 0.24 \\
$A_V^{host}$ (mag) & 1.34 & 0.03 & 0.98 & 0.02 \\
$\Delta m_{15}$ (mag) & 1.04 & 0.04 & 1.06 & 0.03\\
$\mu_0$ (mag) & 27.99 & 0.02 & 28.01 & 0.05\\
$\chi^2$/d.o.f. & 2.99 & & 17.13 & \\  
\hline
%\multicolumn{5}{c}{SALT2.4}\\
SALT2.4 & \multicolumn{2}{c}{$BVRI$} & \multicolumn{2}{c}{$gri$} \\
\hline
$T_{max}$ (MJD) & 56690.88 & 0.11 & 56691.49 & 0.11 \\
$C$ & 1.22 & 0.03 & 0.86 & 0.03 \\
$x_0$ & 0.416 & 0.013 & 0.544 & 0.017 \\
$x_1$ & 0.03 & 0.06 & 1.39 & 0.07 \\
$m_B$ (mag) & 11.50 & 0.03 & 11.25 & 0.03 \\
$\mu_0$ (mag) & 26.74 & 0.13 & 27.80 & 0.12 \\
$\chi^2$/d.o.f. & 5.83 & & 5.87 & \\  
\hline
\hline
\end{tabular}
\end{table}

As seen in Table~\ref{tab-14J-par}, the LCs of SN~2014J have been fit with two
different models assuming different reddening laws for the host galaxy in MLCS2k2 and SNooPy2. 
This was motivated by the fact that many studies (see below) found $R_V < 2$ in M82, 
quite different from the Milky Way value of $R_V = 3.1$. 
  
When using MLCS2k2, we considered two different scenarios: first, we adopted 
$A_V^{MW} = 0.43$ mag from the extinction map of \citet{sf11} at the position 
of SN~2014J, and $R_V = 1.4$ based on the results of 
\citet{goobar14}, \citet{foley14} and \citet{ama14}. 
Secondly, we let $R_V$ float until the lowest $\chi^2$ was found by MLCS2k2.
This resulted in $R_V \sim 1.0$, and the parameters corresponding to this
solution are adopted as the best-fit MLCS2k2 values (see Table~\ref{tab-14J-par}). 
Note that such a low value of $R_V$ is close to the limiting case of Rayleigh
scattering from very small particles, producing $R_V \sim 1.2$ \citep{draine03}.
From the full sample of the SDSS-II SN survey (361 SNe Ia) \citet{lampeitl10}
found that the average extinction law for SNe in passive host galaxies is
$R_V = 1.0 \pm 0.2$. Thus, even though the host of SN~2014J, M82, is an extremely
active star-forming galaxy, such a low value for the extinction law is not
unprecedented. 

In SNooPy2, different reddening laws are implemented as different ``calibrations" 
\citep{burns14}. We applied both the {\tt calibration=3} and {\tt calibration=6} 
settings (corresponding to $R_V \sim 1.5$ and $R_V \sim 1.0$, respectively), and
list the best-fit parameters for each in Table~\ref{tab-14J-par}. 

Note that \citet{marion15} adopted a different Milky Way extinction value toward
SN~2014J (they used $E(B-V)_{MW} = 0.05$ mag corresponding to $A_V^{MW} = 0.16$ mag),
because the dust content of M82 may influence the far-IR maps of \citet{sf11} in that direction. 
Using this lower Milky Way extinction parameter one would get $\sim 0.1$ mag higher
$A_V^{host}$ and $\sim 0.15$ mag higher distance modulus for SN~2014J. While keeping
this in mind, in the following we use the higher Milky Way extinction value
as given by the reddening maps of \citet{sf11}. In this case the MLCS2k2
results are directly comparable to the ones derived by SNooPy2, because
SNooPy2 automatically applies the \citet{sf11} values for calculating the Milky Way extinction.
 
SALT2.4 fits the reddening of the SN in a different way: instead of applying the
same dust extinction law as MLCS2k2 or SNooPy2, it models the reddening via the 
$C$ color parameter (see Eq.\ref{eq3}). Thus, the effect of the strong 
interstellar extinction on the LCs of SN~2014J is reflected by the extremely
large value of its SALT2.4 color coefficient, which is more than an order
of magnitude higher than for the other three SNe. 

The light curves are plotted in Fig.~\ref{fig-14J}.
%\ref{fig-14J-snpy} and \ref{fig-14J-salt2}.

Applying SNooPy2 on their own $UBVRIJHK$ photometry, \citet{marion15} 
obtained $T_{max}(B) = 56689.74 \pm 0.13$ MJD, 
$dm_{15}(B) = 1.11 \pm 0.02$, $E(B-V)_{host} = 1.23 \pm 0.01$
and $\mu_0 = 27.85 \pm 0.09$ mag in {\tt calibration=4} mode 
($R_V \sim 1.46$). 
Their reddening value, $E(B-V)_{host}$, implies $A_V^{host} = 1.80 \pm 0.03$ mag.
These parameters are marginally consistent with our results listed
in Table~\ref{tab-14J-par}. However, the relatively large differences
between the distance moduli obtained from the $BVRI$ and $griz$
data, and also between the results from MLCS2k2 and SNooPy2, suggest
that the $R_V \sim 1.4$ solution may not be the best one as far
as LC fitting is concerned. Indeed, by comparing the distance moduli
obtained from the $R_V \sim 1.0$ solutions, it seems that 
those values are more consistent with each other. The average 
$\mu_0$ for the latter solution is $\sim 28.09 \pm 0.11$ mag, 
while it is $\sim 27.72 \pm 0.20$ mag for the $R_V \sim 1.4$
solution. The dispersion of the distance moduli derived from 
the two sets of light curves and two independent codes is 
much less when $R_V = 1.0$ is used for the M82 reddening law, compared
to the $R_V \sim 1.4$ case adopted by \citet{marion15}.

Fitting the combined $BVRI$ + $g'r'i'$ dataset with SNooPy2 gave
distance moduli similar to those listed in Table~\ref{tab-14J-par}: 
$\mu_0 = 27.50 \pm 0.02$ ({\tt calibration=3}) and $\mu_0 = 28.02 \pm 0.02$
({\tt calibration=6}). However, the reduced $\chi^2$ values for the
combined fits were $31.66$ and $34.79$, respectively, indicating 
poor fitting quality. Comparing the best-fit template LCs with the
observed ones revealed that the $g'$ band data could not be
fit simultaneously with the other bands: while the shape of the template LC 
was similar to the observed one, the observed $g'$-band LC was too bright
(by $\sim 0.5$ mag) with respect to the template. This could be due to either
an issue with our photometry (which is unlikely given that the other data do
not show such a high deviation), or the complexity of the reddening law
in M82 that may not be fully modeled by a single $R_V$ \citep{foley14}.
 
The SALT2.4 code could not provide reliable distances for this
heavily reddened SN. The SALT2.4 distances for SN~2014J 
are inconsistent with each other, as well as with
the distances given by the other two codes. 
SALT2.4 also failed to give consistent fits to the combined 
$BVRI$+$g'r'i'$ LCs: neither of the templates matched the observed
data adequately, resulting in $\chi^2 > 50$. 

We conclude that for SN~2014J only MLCS2k2 and SNooPy2 were able to provide
more-or-less consistent distances, and both of those LC fitters suggest 
$R_V \sim 1.0$, i.e. a lower value than found by the
spectroscopic studies. However, the failure of the simultaneous fitting
of the combined LCs suggest that the complexity of the extinction within M82
may affect the derived distances to SN~2014J more than in the other three cases.

\subsection{Correction for the host galaxy mass}

\begin{table*}
\centering
\caption{Differences in distance moduli after corrections for host galaxy masses}
\label{tab-massstep}
\begin{tabular}{lcccc}
\hline
\hline
SN & SALT$-$MLCS  & SALT$-$MLCS  &  SALT$-$SNooPy  & SALT$-$SNooPy  \\
 & $BVRI$ (mag) & $g'r'i'z'$ (mag) & $BVRI$ (mag) & $g'r'i'$ (mag) \\
\hline 
2012cg & 0.01 ($-0.02$) & -- & 0.11 (0.08) & -- \\
2012ht & $-0.01$ ($-0.04$) & 0.24 (0.21) & $-0.11$ ($-0.14$) & $-0.09$ ($-0.12$) \\
2013dy & 0.04 (0.01) & 0.1 (0.07) & 0.03 (0.00) & 0.31 (0.28) \\
\hline
\hline
\end{tabular} 
\tablecomments{Differences between the uncorrected distance moduli are given in parentheses.}
\end{table*}

The distance moduli given in Tables~\ref{tab-12cg-par}-\ref{tab-14J-par} do not
contain the correction for the host galaxy mass (see Section~\ref{intro} for
references). \citet{betou14} found that SNe Ia that exploded in host galaxies
having total stellar mass of $M_{stellar} > 10^{10}$ $M_\odot$ are
$\sim 0.06$ mag brighter than those in less massive hosts. The calibration of
SALT2.4 by \citet{betou14} that we applied in this paper already contains 
this so-called ``mass-step": the $M_B$ parameter given after Eq.\ref{eq4} is valid
for SNe in less massive hosts, and it is $M_B - 0.061$ mag for SNe in more
massive hosts. 

The stellar masses for the host galaxies in this paper are listed in Table~\ref{tab-basic}.
These were derived, following \citet{pan14}, by applying 
Z-PEG\footnote{\tt http://imacdlb.iap.fr/cgi-bin/zpeg/zpeg.pl} \citep{leborgne02} 
to the observed galaxy SEDs (see Table~\ref{tab-basic} for references).  
It is seen that only SN~2014J is affected by this correction, since the hosts of
the other three SNe are below $\log M_{stellar} = 10$. Thus, their $M_B$ parameter
does not need to be corrected, and their SALT2.4 distances are final. 
Unfortunately, the SALT2.4 distance moduli of SN~2014J are unreliable as they are 
affected by the strong non-standard interstellar extinction
(see the previous subsection). These systematic uncertainties 
($\gtrsim 0.5$ mag; Table~\ref{tab-14J-par}) are much higher 
than the correction for the host galaxy mass ($\sim 0.06$ mag) in the case
of SN~2014J.

Nevertheless, we investigated whether the mass-step correction could bring
the derived distances to better agreement with each other for the other three SNe.
Since neither MLCS2k2, nor SNooPy2 contain the mass-step correction in their calibrations,
we followed the practice applied by \citet{riess16} by adding $0.03$ mag to the 
MLCS2k2/SNooPy2 distance moduli of SN~2014J and subtracting $0.03$ mag from the
distances of SNe~2012cg, 2012ht and 2013dy, thus, mimicking the existence of the
mass-step in their peak brightnesses. 

Table~\ref{tab-massstep} shows the differences between the distance moduli estimated by
SALT2.4 and the other two codes for both photometric systems after implementing the
host mass correction as described above. For comparison, we give the same differences between
the uncorrected distances in parentheses. Ideally, after correction all these differences
should be zero. In reality, it is apparent that the effect of the host mass correction is 
minimal: sometimes it makes the agreement slightly better, sometimes slightly worse, but 
its amount ($\pm 0.03$ mag) is an order of magnitude less than the differences between
the distance moduli given by the different LC-fitting codes. 

It is concluded that the host mass correction on the distance modulus is negligible compared
to the other sources of uncertainty, at least for the SNe studied in this paper. Note, however,
that in studies using {\it relative} distances, such as \citet{betou14,riess16} and others,
this effect can be much more important and significant. Therefore, 
in the rest of the paper we use the uncorrected distances (i.e. without the mass-step) as given
in Tables~\ref{tab-12cg-par}-\ref{tab-14J-par}, but note that it is desirable to include the 
host mass correction in future retraining of the MLCS2k2 and/or SNooPy2 templates.     

\section{Discussion}\label{disc}

In this section we cross-compare the parameters derived by the three independent
codes, and check the consistency between the values inferred from different 
photometric systems ($BVRI$ vs $griz$) and by different LC fitters. 

\subsection{Time of maximum light}

For Type Ia SNe the moment of maximum light in the $B$-band has been used 
traditionally as the zero-point of time. At first it seems to be fairly easy
to measure directly from the data, at least when the LC in $B$-band is available. 
From Tables~\ref{tab-12cg-par} - \ref{tab-14J-par} it is seen that the 
LC fitters used in this study do a good job in estimating $T_{max}(B)$ even
if the $B$-band LC is not included in the fitting. The consistency 
between the derived values is also relatively good: the dispersion around
the mean values is 0.49, 0.21, 0.71 and 0.67 day for SN~2012cg, 2012ht, 2013dy
and 2014J, respectively. Note, however, that SALT2.4 gets later maximum times 
systematically by $\Delta t > 0.5$ day relative to MLCS2k2, similar to 
the finding by \citet{vinko12} and \citet{pere13} for SN~2011fe. 

\subsection{Extinction}

\begin{figure*}
\centering
\resizebox{8cm}{!}{\includegraphics{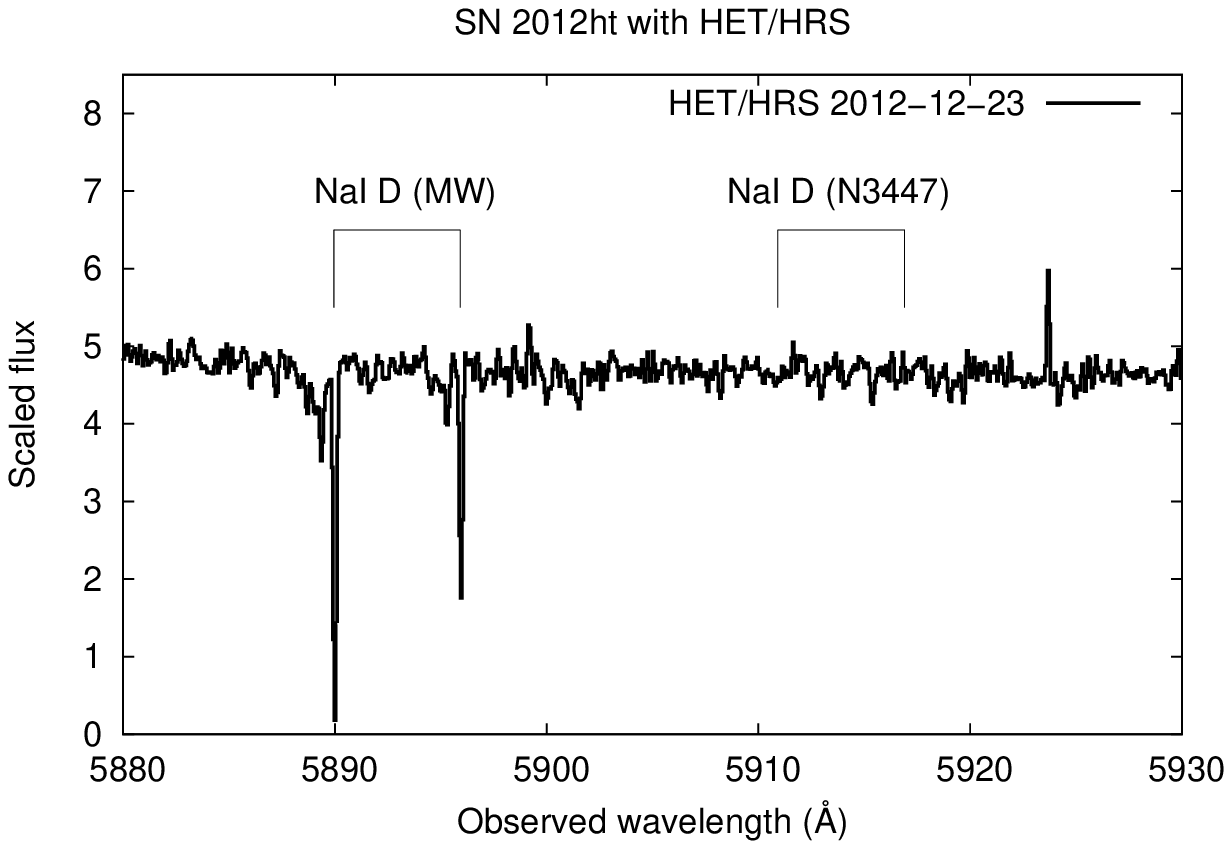}}
\resizebox{8cm}{!}{\includegraphics{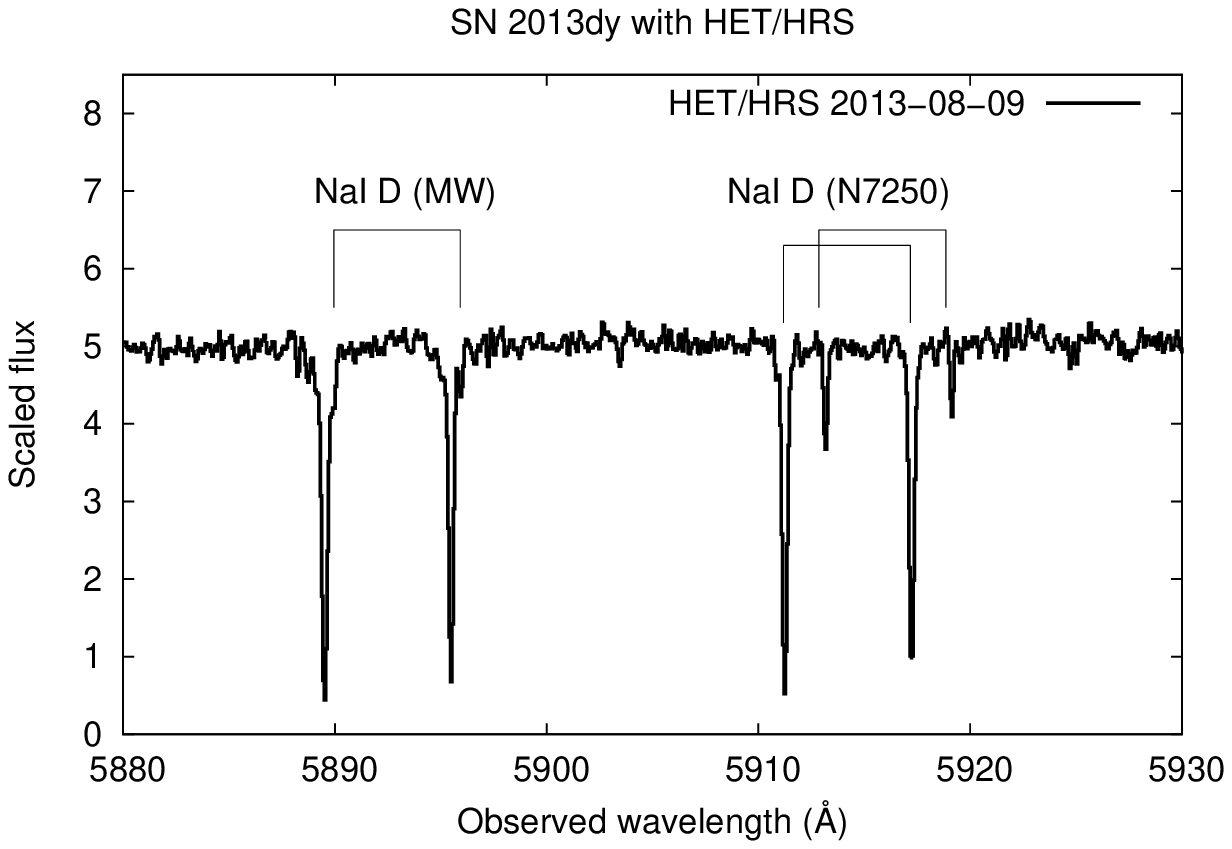}}
\caption{The narrow NaD features in the high-resolution spectrum of SN~2012ht taken at $11$ days before maximum 
(left panel) and 2013dy at $12$ days after maximum (right panel).
The components from the Milky Way and host galaxy are labeled. See text for details.}
\label{fig-nad}
\end{figure*}

The host galaxy dust extinction parameters ($A_V^{host}$) in Tables~\ref{tab-12cg-par} - \ref{tab-14J-par}
look generally consistent with each other. The match between the values provided by
the same code for $BVRI$ and $griz$ is usually better than the agreement between 
the results of the different codes (here only MLCS2k2 and SNooPy2 are relevant, because
SALT2.4 does not model dust extinction). 

As an independent sanity check, we compare the average $A_V^{host}$ values from MLCS2k2
and SNooPy2 for SN~2012ht and 2013dy with high-resolution spectra obtained with the
Hobby-Eberly Telescope (HET). For SN~2013dy these spectra were published by \citet{pan15},
while those for SN~2012ht are yet unpublished. In Fig.~\ref{fig-nad} the spectral regions
containing the Na D ($\lambda\lambda$5890,5896) doublet are shown. The narrow Na D 
features originating from the ISM both in the Milky Way and in the host galaxy (separated
by the Doppler-shift due to the recession velocity of the host) are labeled.
The strength of the Na D doublet is thought to be proportional to the amount 
of extinction, at least as a first approximation \citep{rich94, poz12}.
It is seen that the dust extinction within the host galaxy for SN~2012ht is 
negligible (no narrow Na D absorption is visible at the redshift of the host) 
compared to the Milky Way component. This is in excellent agreement with the predictions from
the LC fitters, because both codes resulted in $A_V^{host} = 0$ magnitude for SN~2012ht.

Concerning SN~2013dy, the consistency between the photometric and spectroscopic extinction
estimates is also very good. In Fig.~\ref{fig-nad} the Na D profiles in the host galaxy
have approximately the same strength as the Milky Way components. From 
Tables~\ref{tab-basic} and \ref{tab-13dy-par}, $A_V^{MW} = 0.42$ mag and 
$A_V^{host} \sim 0.44 \pm 0.14$ mag were taken for SN~2013dy, which are, again, 
in good agreement with the relative strengths of the Na~D profiles in the high-resolution spectra. 

The case of SN~2014J is more problematic, as this SN occured within a host galaxy that has
a known complex dust content. \citet{foley14} presented an in-depth study of the 
wavelength-dependent reddening and extinction toward SN~2014J, and concluded that it is 
probably much more complex than a simple extinction law parametrized by a single value of
$A_V$ and $R_V$. Keeping this in mind, it is not surprising that the LC-fitting codes
applied in this paper failed to produce consistent results with the spectroscopic estimates
\citep{ama14, foley14, goobar14, brown15, gao15} that all found $R_V \gtrsim 1.4$.

\subsection{Light curve shape/stretch parameter}

\begin{table}
\centering
\caption{}
\label{tab-stretch}
\begin{tabular}{lcccc}
\hline
\hline
SN & $\Delta m_{15}(B)$ & $\Delta m_{15}(B)$ & $\Delta m_{15}(B)$  \\
 & $BVRI$ (mag) & $g'r'i'z'$ (mag) & combined (mag) \\
\hline 
2012cg & 0.984 (0.041) &  -- & --  \\
2012ht & 1.275 (0.045) & 1.217 (0.041) & 1.298 (0.010) \\
2013dy & 0.960 (0.043) & 0.858 (0.015) & 0.971 (0.007) \\
2014J  & 1.034 (0.065) & 0.952 (0.105) & 0.981 (0.074) \\
\hline
\hline
\end{tabular} 
\tablecomments{Standard deviations are given in parentheses.}
\end{table}

The light curve shape/stretch parameter is the one that is most strongly connected to the peak brightness of
a Type Ia SN; thus, it has a direct influence on the distance measurement. Since the three
LC-fitting codes adopt slightly different parametrizations of the light curve shape, we
converted each of them to the traditional $\Delta m_{15}(B)$ \citep{phil93}. From the
MLCS2k2 templates we get $\Delta m_{15}(B) ~=~ 1.07 ~+~ 0.67 \cdot \Delta ~-~0.10 \cdot \Delta^2$.
For SNooPy2 we adopted $\Delta m_{15}(B) ~=~ 0.13 ~+~ 0.89 \cdot \Delta m_{15}$ \citep{burns11},
while for SALT2.4 we used 
$\Delta m_{15}(B) ~=~ 1.09 - 0.161 \cdot x_1 + 0.013 \cdot x_1^2 - 0.0013 \cdot x_1^3$ 
\citep{guy07}.  

Table~\ref{tab-stretch} lists the resulting $\Delta m_{15}(B)$ values, averaged over the three methods,
for the $BVRI$ and $g'r'i'z'$ data and for the combined fits, respectively. It is seen that the fits
to the $g'r'i'z'$ data alone tend to result in systematically lower decline rates than the fits to
the $BVRI$ data or the combined $BVRI$+$g'r'i'$ data. The deviation is the highest in the case of 
SN~2013dy, $\sim 0.1$ mag, and it is lower for the other two SNe. Neglecting the differences 
between the other fitting parameters, an underestimate of the decline rate by $\sim 0.1$ mag 
would cause an overestimate of $\sim 0.08$ mag in the distance modulus 
(applying the decline rate - absolute peak magnitude calibration by \citet{burns14}). 
Since the derived distance moduli do not show such a systematic trend between the $BVRI$ 
and $g'r'i'z'$ data, and  the differences between them are sometimes higher 
than 0.08 mag (see next section), it is concluded that the systematic underestimate 
of the decline rates from our $g'r'i'z'$ photometry is not significant,
at least from the present dataset. The number of SNe in this paper is too low to draw more definite
conclusions on the cause of the dependence of the light curve shape/stretch parameter
on the photometric bands (whether it is merely due to uncertainties in the data or might have physical origin), but it would be worth studying on a larger sample of SNe.

\subsection{Distance}

\begin{figure*}
\centering
\resizebox{8cm}{!}{\includegraphics{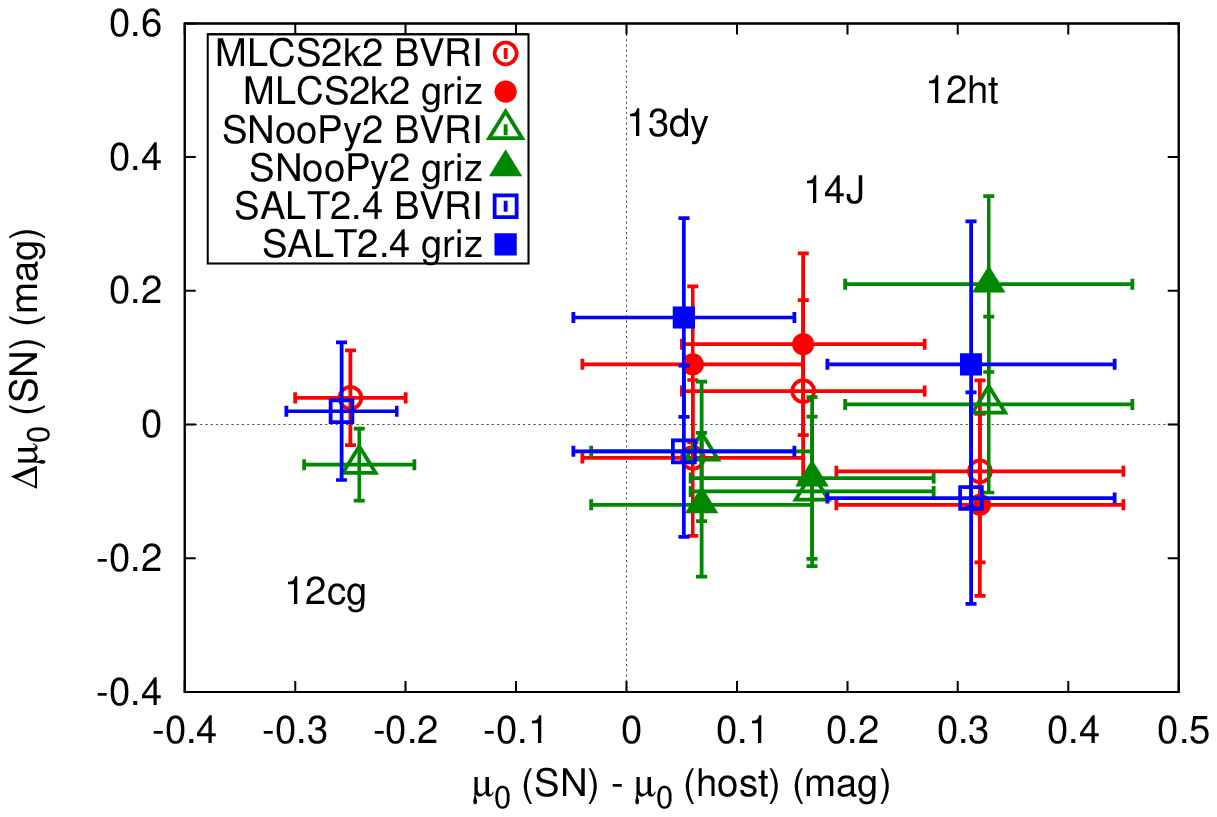}}
\resizebox{8cm}{!}{\includegraphics{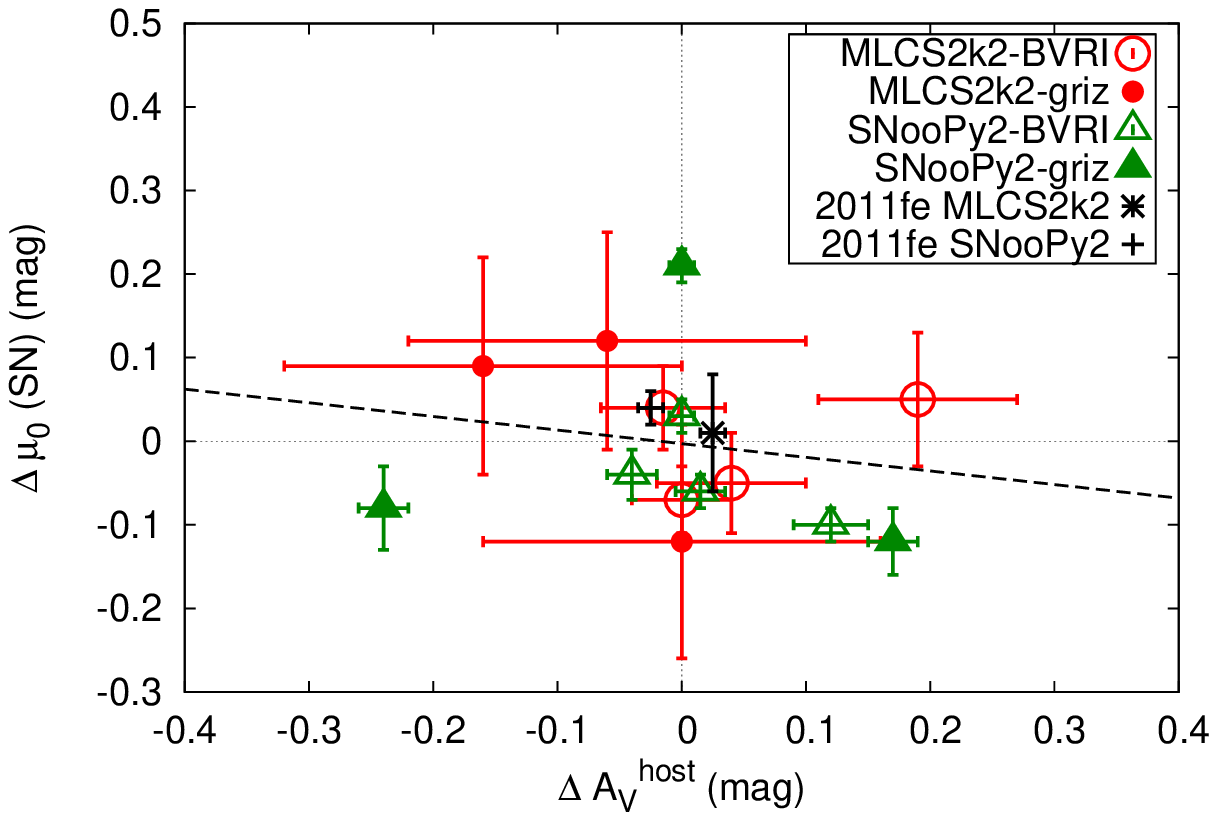}}
\caption{Left panel: Residual distance moduli from their mean value for each SNe. 
Different symbols denote the different LC-fitting codes, as explained in the legend. 
Right panel: the distance modulus residuals as a function of the dust extinction residuals. 
The dashed line indicates a simple linear fit to these data. See text for explanation.}
\label{fig-mu0}
\end{figure*}

\begin{table*}
\centering
\caption{Final absolute distances to SNe Ia}
\label{tab-finaldist}
\begin{tabular}{lccccc}
\hline
\hline
SN & $\mu_0$(SN) & $\mu_0$(host) & $D_{SN}$ & $D_{host}$ & Difference \\
   & (mag) & (mag) & (Mpc) & (Mpc) & (mag)\\
\hline
2011fe & 29.13 $\pm 0.08$ & 29.13 $\pm 0.04$ & $6.7^{+0.3}_{-0.2}$ & $6.7^{0.1}_{0.1}$ & 0.00 ($< 1 \sigma$)\\
2012cg & 30.83 $\pm$0.05 & 31.08 $\pm$0.29 & $14.6^{+0.3}_{-0.3}$ & $16.44^{+2.4}_{-2.1}$ & $-0.25$ ($\lesssim 1 \sigma$)\\
2012ht & 32.23 $\pm$0.13 & 31.91 $\pm$0.04 & $27.9^{+1.7}_{-1.6}$ & $24.1^{+0.5}_{-0.5}$ & 0.32 ($\sim 2.3 \sigma$)\\
2013dy & 31.54 $\pm$0.08 & 31.50 $\pm$0.08 & $20.3^{+0.8}_{-0.7}$ & $20.0^{+0.8}_{-0.7}$ & 0.04 ($\lesssim 1 \sigma$)\\
2014J & 28.09 $\pm$0.11 & 27.93 $\pm$0.35 & $4.1^{+0.2}_{-0.2}$ & $3.9^{+0.7}_{-0.6}$ & 0.16 ($\lesssim 1 \sigma$)\\ 
\hline
\end{tabular}
\end{table*}   

Distance is one of the most important outputs of the LC fitting codes under study. Cross-comparing
the distance estimates produced by the various codes may reveal important constraints on the
systematics that are present either in the basic assumptions of the methods or in their
implementation and calibrations. 

For the extremely well-observed SN~2011fe, \citet{vinko12} found a $0.16 \pm 0.07$ mag systematic 
difference ($\sim 2 \sigma$) between the distance moduli provided by MLCS2k2 and SALT2,
when applied for the same homogeneous photometric data. 
In the left panel of Fig.~\ref{fig-mu0} we
plot the residual distance moduli (i.e. the difference of the distance modulus given by the
LC-fitting code for each filter set, $BVRI$ or $griz$, from the mean distance modulus) for each
SNe as a function of the difference between their mean distance moduli and the ones listed in Table~\ref{tab-basic}. 
The results from fits to $BVRI$ data are plotted with open symbols, while those from $griz$ data
are shown by the filled symbols. The color and the shape of the symbols encode the fitting method as indicated by the figure legend.

From this plot it is seen that a separation of $\Delta \mu_0 \sim 0.2$ mag can still be 
present between the distance moduli given by different LC-fitting codes, similar to the case of SN~2011fe,
although the difference varies from SN to SN and it is less than $\sim 0.2$ mag for the majority of
the cases considered in the present paper.  

For the three moderately reddened SNe (2012cg, 2012ht and 2013dy) the distances given by MLCS2k2 and SALT2.4
are in remarkable agreement (their difference is $0.02$, $0.04$ and $-0.01$ mag, respectively) for the $BVRI$ data.
The differences between the MLCS2k2 and SNooPy2 distances in $BVRI$ are also very good; they do not exceed 
$0.1$ mag. This is not true in the case of SN~2014J, as expected, because SALT2.4 could not provide reliable
distances for such an extremely reddened object, thus, they are not plotted in Fig.~\ref{fig-mu0}.
Note that this does not mean that SALT2.4 is a less
reliable code. It is just the consequence of the underlying model, and the code works fine for
moderately reddened SNe Ia. 

The agreement is slightly worse for our $g'r'i'z'$ photometry, partly because those data have smaller 
signal-to-noise ratio than our $BVRI$ light curves. From these data it is found that the 
differences between the distance moduli 
provided by the three codes may differ by $\sim 0.2$ mag. A similar $\lesssim 0.2$ mag difference 
can be seen when comparing the distances of the same SN taken from $BVRI$ and $g'r'i'z'$ LCs:
for SN~2012ht $\mu_0(BVRI) - \mu_0(g'r'i'z')$ is $0.05$, $-0.18$ and 
$-0.30$ mag from MLCS2k2, SNooPy2 and SALT2.4, respectively; for SN~2013dy these are $-0.14$,
$0.08$ and $-0.20$; for SN~2014J $-0.07$, $-0.02$ and $-1.06$ mag. It is seen that there is
no systematic trend in these numbers, which suggests that these differences are probably not
due to systematic errors in the photometric calibration; more probably they represent the
internal uncertainties of the template LC vectors for the different bands.
Given that our data were obtained by
two telescopes from two different sites (one for $BVRI$ and another one for $g'r'i'z'$), this result may also 
give a hint on the possible amount of systematic errors in the distance moduli when fitting 
inhomogeneous LCs taken by more than one telescope.

It is concluded that applying these three popular LC-fitting codes as 
distance calculators on homogeneous $BVRI$ photometry on nearby, moderately 
reddened SNe Ia one can derive consistent distances that agree with each other within
 $\lesssim 0.1$ mag or better. Even for strongly
reddened SNe, like SN 2014J, MLCS2k2 and SNooPy2 work quite well; their distances differ by only 
$\sim 0.15$ mag. We found slightly larger ($\sim 0.2$ mag) differences in the distances 
from our $g'r'i'z'$ photometry. 

In the right panel of Fig.~\ref{fig-mu0} we investigate whether the differences between the distance
moduli were due to systematic under- or overestimates of the extinction parameter $A_V^{\mathrm host}$. 
In this panel the same $\Delta \mu_0$ residuals are plotted against the residual of the host 
extinction parameter $\Delta A_V^{\mathrm host}$ (we consider only the extinction within the host here, since the Milky Way
extinction was kept fixed at the values provided by the Milky Way dust maps). SN~2011fe is also plotted
in this diagram (with black symbols) taking the data from \citet{vinko12}. 
Again, only the MLCS2k2 and SNooPy2 results are used.

If the expected correlation between the distances and extinction (higher extinction estimates imply
shorter distances) exists, then one should see positive $\Delta \mu_0$ values for negative $\Delta A_v^{host}$
and vice versa. It is not clearly visible in Fig.~\ref{fig-mu0}, as most of the data scatter around 0 within
$\sim 0.2$ mag in all directions. 
The dashed line is a simple linear fit to the data. Its slope, $-0.16 \pm 0.23$, 
is in the expected direction of a correlation between the extinction and distance, 
though it is not significant. Although it is
probable that our sample is too low to detect this, the lack of significant correlation 
suggests that the distances given by either MLCS2k2 or SNooPy2 are less affected by systematic under- or overestimates of the extinction parameter.

In Table~\ref{tab-finaldist} we summarize the final absolute distances for each SN studied, 
supplemented by the data for SN~2011fe from \citet{vinko12}.   
These are defined as the simple, unweighted mean of all distance moduli from our Konkoly + Baja
data except for SN~2014J where the SALT2.4 distances were omitted due to the reason mentioned above.
The Cepheid-based distances to the host galaxies \citep{riess16} from Table~\ref{tab-basic} are
also shown for easy comparison. In the final column the difference between the SN and host distance
moduli is given with respect to their combined uncertainties 
$\sigma(\mu_0) = \sqrt{\sigma^2_{SN} + \sigma^2_{host}}$. It is apparent that even though the individual
SN-based mean distances are uncertain at the $\sim 0.10$ mag level, their deviations from their
host galaxy distances obtained independently are less than $1 \sigma$ in 4 out of 5 cases, which is
encouraging. Higher deviation ($2 \sigma$) is seen only in the case of SN~2012ht. 

Recently \citet{riess16} showed that by properly combining Cepheid- and SN~Ia-based distance scales,
anchored to local galaxies having independent geometric distances, one can reduce the uncertainty of
the local value of the Hubble-constant ($H_0$) to $\sim 2.4$ percent. Such an accuracy on the 
{\it individual} distances to local galaxies would need a $\sim 0.05$ mag dispersion in the distance 
moduli. Our results above show that this is still not the case for every SN at present, although
the agreement between the distance moduli are close to the $\lesssim0.05$ mag level for the 
best-observed SNe in our sample. The $\sim 0.1$ -- 0.2 mag
dispersion between the distance moduli calculated by different LC fitters is close to 
the $\sim 0.15$ mag dispersion found by \citet{riess16} when comparing their Cepheid- 
and SN~Ia-based distances to the same galaxies.

As the sample of the SNe~Ia having accurately calibrated
photometry \citep{scolnic15} is growing rapidly, we can expect significant improvement in the accuracy of
the individual distance estimates to local galaxies in the near future. This would be an important step
toward better understanding the physics of SN Ia explosions.

\section{Summary}\label{sum}

We have studied 3 public light curve fitting codes for SNe Ia by cross-comparing the time of maximum,
extinction and distance parameters inferred from the light curves of 4 nearby, bright, well-observed
SNe Ia (2012cg, 2012ht, 2013dy, 2014J). Our results are summarized as follows.
\begin{itemize}
\item{The moment of $B$-band maximum can be estimated within $\pm 0.7$ day, even if
there are no data observed in the $B$-band. Note that SALT2.4 tends to give systematically later
maximum times by $\Delta t_{max} \sim 0.5$ day than MLCS2k2 \citep{vinko12, pere13}.}

\item{For moderately reddened ($A_V^{host} < 0.5$ mag) SNe, MLCS2k2 and SNooPy2 did quite a good
job in estimating the relative amount of interstellar extinction within the host galaxy ($A_V^{host}$)
compared to the extinction within the Milky Way ($A_V^{MW}$).  The inferred $A_V^{host} / A_V^{MW}$
extinction ratios for SNe 2012ht and 2013dy are consistent with the relative strengths of the 
interstellar NaD lines from high-resolution spectroscopy. 
This is not the case for the heavily-reddened SN~2014J, where 
the light curve fitting resulted in reddening parameters having large scatter and being different
from the results of spectroscopic analyses. }

\item{Regarding the distance modulus, it is found that for the moderately reddened SNe 2012cg, 2012ht 
and 2013dy the consistency between the results from the three LC-fitting codes is $\lesssim 0.1$ mag 
for our highest quality $BVRI$ data, even without taking into account the dependence of the peak
brightnesses on the host galaxy masses. 
This is significantly better than the $\sim 0.16$ mag difference found by \citet{vinko12} for SN~2011fe.
The dispersion is somewhat higher, $\sim 0.2$ mag, for our $g'r'i'z'$ LCs that have lower S/N ratio, 
and the same is true for the strongly reddened SN~2014J. 
We found a negative, though nonsignificant, distance--extinction correlation in our sample, 
suggesting that the distances provided by both MLCS2k2 and SNooPy2 are not strongly 
affected by systematic over- or underestimates of the extinction.
The final distances to our sample SNe are in very good agreement with the Cepheid-based distances
to their host galaxies (see Table~\ref{tab-finaldist}). From these results it seems important to 
utilize only low-extinction ($A_V < 0.5$ mag) SNe Ia to measure $H_0$ in order to reduce the 
potential systematic errors on the derived distances due to dust extinction \citep{riess16}.}
  
\end{itemize}

\begin{acknowledgements}
We are indebted to Adam Riess for his valuable comments and suggestions on the previous version of this
paper. We also thank the thorough work of an anonymous referee for his/her report that helped
significantly to improve the paper.
This work is part of the project ``Transient Astrophysical Objects" GINOP 2.3.2-15-2016-00033 of the 
National Research, Development and Innovation Office (NKFIH), Hungary, funded by the European Union.  
Some co-authors of this publication have also received funds from the following grants: 
Hungarian NKFIH/OTKA PD-116175 (LM), PD-112325 (TS), K-109276 (KV), K-113117 (KV) and NN-107637 (JV); MTA-Lendulet LP2012-31 (AP); NSF Fellowship AST-1302771 (JMS); NSF Grant AST-1109881 (JCW). 
JCW is supported by the Samuel T. and Fern Yanagisawa Regents Professorship.
LM and KV are supported by the Bolyai Janos Research Scholarship of the Hungarian Academy of Sciences.

\end{acknowledgements}

%\end{document}
\newpage

\begin{appendix}

\section{The construction of $g'r'i'z'$ vectors for MLCS2k2}

\begin{figure*}
\centering
\resizebox{8cm}{!}{\includegraphics{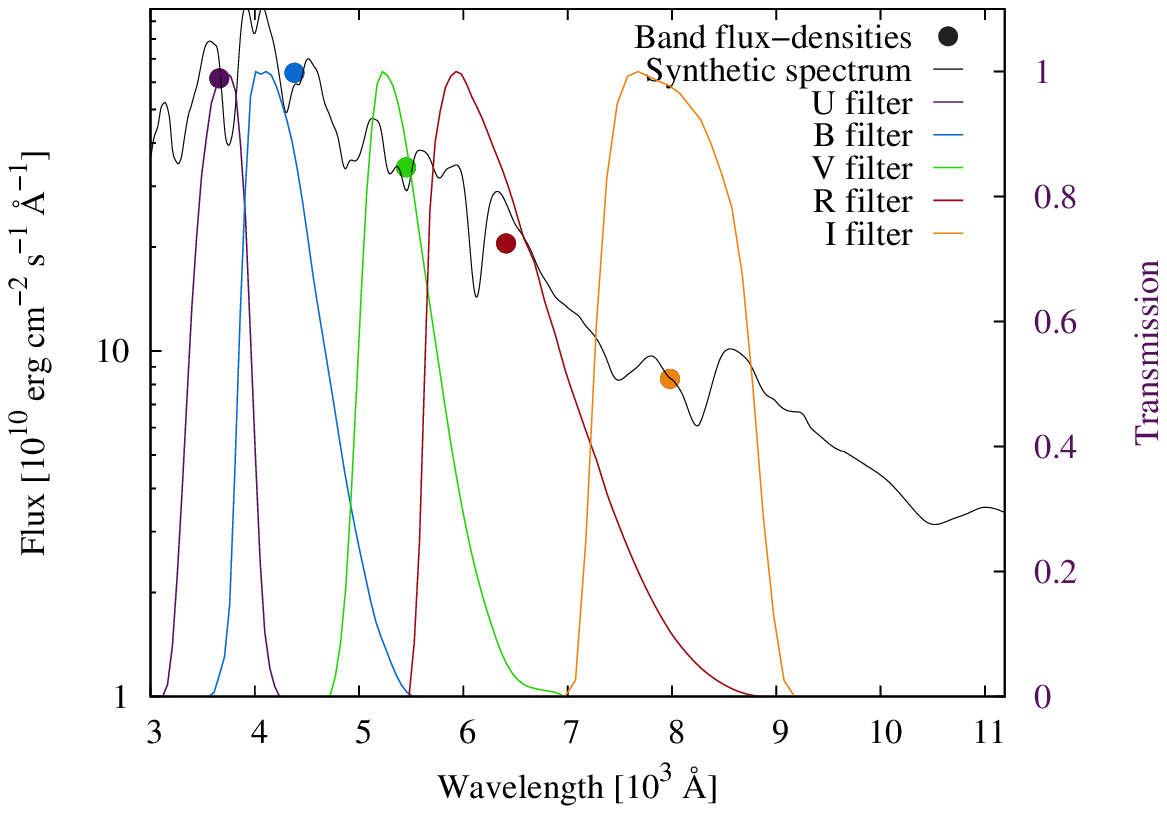}}
\resizebox{8cm}{!}{\includegraphics{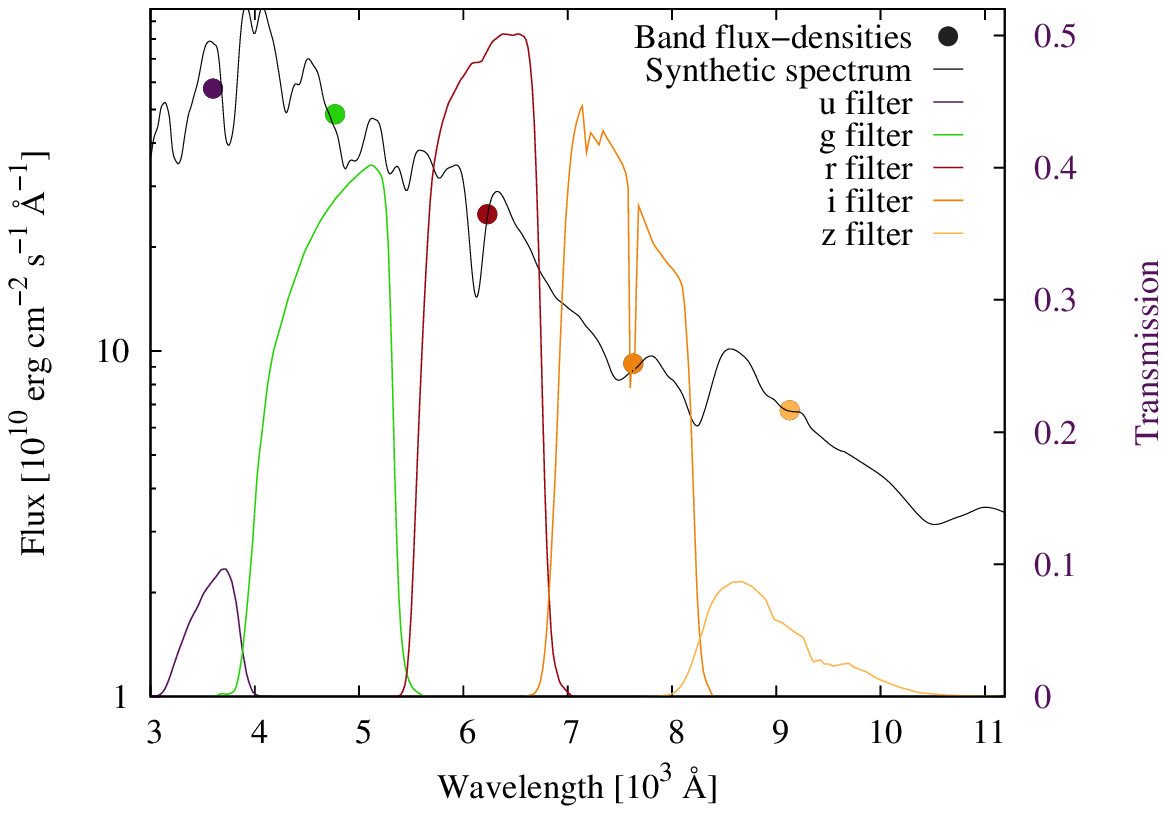}}
\caption{Left panel: The Hsiao template spectrum at maximum light and the transmission curves of the 
Johnson-Cousins $UBVRI$ system. Filled symbols indicate the synthetic flux densities. Right panel: the
same but for the Sloan $u'g'r'i'z'$ system.}
\label{fig:hsiao}
\end{figure*}

Since MLCS2k2 is originally designed for only the $UBVRI$ system \citep{jrk07}, one needs to transform 
its $M_X$, $P_X$ and $Q_X$ vectors ($X = \{U,B,V,R,I\}$) to other filters if data from other photometric system
are to be fit. We have computed the transformation to the Sloan $u'g'r'i'z'$ system via the Hsiao 
template spectra \citep{hsiao07} in the following way.

First, synthetic fluxes for both $UBVRI$ and $u'g'r'i'z'$ filters were computed from the Hsiao template spectra by convolving the templates with the corresponding filter functions (Fig.\ref{fig:hsiao}). This was done for all
templates between $-10$ day and $+90$ day. Next, the following flux ratios were defined as the basis of the
$UBVRI$ $\rightarrow$ $u'g'r'i'z'$ transformation: $f(u') / f(U)$, $f(g') / f(B)$, $f(r')/f(R)$, $f(i') / f(I)$
and $f(z') / f(I)$. Among many other combinations, these flux ratios were found to exhibit the least amount of
variation in the $[-10$d,+90d$]$ phase interval. Third, using these flux ratios the MLCS2k2 $UBVRI$ magnitudes
were transformed to fluxes, multiplied by the corresponding flux ratios and converted back to Sloan 
$u'g'r'i'z'$ magnitudes. Note that this conversion is based on the implicit assumption that the above flux ratios
remain the same for SNe having different MLCS2k2 $\Delta$ parameter. This certainly breaks down for 
SNe having large ($> 0.5$) $\Delta$; however, since our SNe have only $\Delta$ within $\pm 0.3$, this assumption
should be approximately valid. The feasibility of the whole transformation can also be judged by the consistency
between the MLCS2k2 fitting parameters computed from the quasi-simultaneous $UBVRI$ and $u'g'r'i'z'$ data. 

MLCS2k2 also contains a prescription for the time-dependence of the extinction in the $UBVRI$ bands, parametrized
as $\zeta_X(t) = A_X(t) / A_X(t_{Bmax})$, where $A_X$ is the extinction in the $X$-band. We transformed the
$\zeta_X$ vectors of \citet{jrk07} to the $u'g'r'i'z'$ system using a similar approach as above. 
First, following \citet{jrk07}, 
the Hsiao templates were reddened with a chosen $A_V$ and using the Cardelli-law \citep{cardelli89}:
\begin{equation}
  {A(x) \over A_V} = ({\alpha}(x) + {{\beta}(x) \over R_V}).
\label{eq:cardelli}
\end{equation}
The standard Milky Way reddening slope, $R_V = 3.1$, was assumed at first, but this condition was relaxed
later (see below). 

Next, both the reddened and the reddening-free $UBVRI$ magnitudes were transformed to $u'g'r'i'z'$ magnitudes as
above, and their time-dependent differences were used to construct the $\zeta$ vectors for the Sloan system:
$\zeta_X = (M'_X(t) - M_X(t)) / (M'_X(t_{Bmax}) - M_X(t_{Bmax}))$, where $M'_X$ denotes the reddened MLCS2k2
magnitude in the $X$-band. 

To be able to handle non-standard (i.e. $R_V \neq 3.1$) reddening slopes, the procedure described above has
been repeated for $1 < R_V < 6$. It was found that the resulting $\zeta_X(R_V)$ values as a function of 
$1-\zeta_X(3.1)$ can be approximated by a parabola of the following form:
\begin{equation}
   1-\zeta_X(R_V) = a_X \cdot (1-\zeta_X(3.1))^2 + b_X \cdot (1-\zeta_X(3.1)) + c_X.
\label{eq:zeta}
\end{equation}

Since the $a_X$, $b_X$ and $c_X$ coefficients are different for every $R_V$, their dependency on $R_V$
were determined by fitting a cubic polynomial to these data as a function of $R_V^{-1}$:
%\begin{equation}
\begin{eqnarray}\label{eq:abc}
   a_X &=& d_{1,1} \cdot (R_V^{-1})^3 + d_{1,2} \cdot (R_V^{-1})^2 + d_{1,3} \cdot (R_V^{-1}) + d_{1,4} \nonumber \\
   b_X &=& d_{2,1} \cdot (R_V^{-1})^3 + d_{2,2} \cdot (R_V^{-1})^2 + d_{2,3} \cdot (R_V^{-1}) + d_{2,4} \\
   c_X &=& d_{3,1} \cdot (R_V^{-1})^3 + d_{3,2} \cdot (R_V^{-1})^2 + d_{3,3} \cdot (R_V^{-1}) + d_{3,4} \nonumber
\end{eqnarray}
%\end{equation}

The fit parameters for both photometric systems are collected in Table~\ref{tab:abc}.

\begin{table*}
\caption {The constant parameters from the third-order polynomial fit to 
$a_X$, $b_X$ and $c_X$ as functions of $R_V^{-1}$ (see Eq.\ref{eq:abc}).}
\label{tab:abc}
\centering
\tiny
\begin{tabular}{lcccccccccc}
  \hline
  \hline
  & $U$ & $B$ & $V$ & $R$ & $I$ & $u'$ & $g'$ & $r'$ & $i'$ & $z'$ \\
  \hline
  $d_{1,1}$ & -29.637 & 1.711 & 0.355 & -1.157 & -11.212 & -41.028 & 4.925 & -0.974 & -8.647 & -193.074 \\
  $d_{2,1}$ & 2.790 & 0.845 & 0.023 & 0.235 & 5.962  & 2.828 & 0.424 & 0.120 & -1.468 & 13.272 \\
  $d_{3,1} \cdot 10^{-3}$ & 0.149 & -0.222 & 0.053 & 0.157 & -6.828 & -0.126 & 1.938 & 0.028 & -0.718 & 44.466 \\
  $d_{1,2}$ & 81.694 & -5.761 & -0.618 & -1.431 & 9.372 & 108.431 & -9.990 & -0.590 & 0.074 & 218.568 \\
  $d_{2,2}$ & -7.388 & -2.551 & -0.236 & 0.339 & -4.992  & -5.476 & -1.879 & 0.268 & 6.286 & -15.035 \\
  $d_{3,2} \cdot 10^{-3}$ & -0.443 & 0.777 & -0.676 & 0.215 & 5.714 & -1.256 & -3.057 & 0.051 & -0.621 & -50.453 \\
  $d_{1,3}$ & -77.845 & -0.101 & -2.496 & -3.137 & -6.392 & -79.613 & 5.363 & 6.273 & 1.242 & -91.518 \\
  $d_{2,3}$ & 7.736 & 4.200 & 2.503 & 1.409 & 3.171 & 6.257 & 3.709 & 1.259 & -2.521 & 6.292 \\
  $d_{3,3} \cdot 10^{-3}$ & 0.401 & 0.409 & 1.466 & 1.314 & -3.743 & 0.560 & 0.973 & 0.785 & 0.378 & 21.295 \\
  $d_{1,4}$ & 17.612 & 0.573 & 0.855 & 1.199 & 1.443 & 15.798 & -0.897 & -1.937 & -0.173 & 12.752 \\
  $d_{2,4}$ & -0.822 & -0.118 & 0.216 & 0.502 & 0.306 & -0.545 & -0.013 & 0.562 & 1.256 & 0.124 \\
  $d_{3,4} \cdot 10^{-3}$ & -0.088 & -0.205 & -0.404 & -0.452 & 0.831 & -0.046 & -0.084 & -0.259 & -0.043 & -2.997\\
  \hline
  \hline
\end{tabular}
\end{table*}

\newpage
\section{Photometric data}

\begin{table*}
\caption{Local $BVRI$ standard stars in the vicinity of SN~2012cg}
\label{tab-12cg-std}
\centering
\tiny
\begin{tabular}{cccccccccc}
\hline
\hline
R.A. & Dec. & B & $\sigma_B$ & V & $\sigma_V$ & R & $\sigma_R$ & I & $\sigma_I$ \\
\hline
12:27:28.9 & 9:29:33.6 & 16.241 &  0.011 & 15.648 &  0.013 & 15.345 &  0.011 & 14.932 &  0.017 \\
12:26:48.0 & 9:28:50.5 & 15.408 &  0.011 & 14.644 &  0.013 & 14.245 &  0.011 & 13.771 &  0.017 \\
12:26:48.3 & 9:29:57.3 & 15.805 &  0.011 & 14.998 &  0.013 & 14.570 &  0.011 & 14.050 &  0.017 \\
12:27:15.9 & 9:27:27.0 & 15.046 &  0.011 & 14.506 &  0.013 & 14.232 &  0.011 & 13.841 &  0.017 \\
12:27:29.8 & 9:23:53.2 & 15.919 &  0.011 & 15.078 &  0.013 & 14.633 &  0.011 & 14.106 &  0.017 \\
\hline
\hline
\end{tabular}
\end{table*}

\begin{table*}
\caption{Photometry of SN~2012cg. The $BVRI$ data are given in Vega-magnitudes. Errors
are given in parentheses.}
\label{tab-12cg-phot}
\centering
\tiny
\begin{tabular}{ccccc}
\hline
\hline
MJD & B & V & R & I \\
\hline
56066.8 & 14.995 (0.031) & 14.679 (0.025) & 14.593 (0.023) & 14.420 (0.048) \\
56067.9 & 14.493 (0.014) & 14.221 (0.010) & 14.133 (0.011) & 13.974 (0.018) \\
56069.8 & 13.629 (0.014) & 13.490 (0.010) & 13.375 (0.009) & 13.229 (0.018) \\
56072.9 & 12.765 (0.008) & 12.674 (0.010) & 12.575 (0.008) & 12.462 (0.017) \\
56073.8 & 12.646 (0.015) & 12.475 (0.016) & 12.481 (0.012) & 12.360 (0.019) \\
56075.8 & 12.452 (0.012) & 12.175 (0.012) & 12.233 (0.007) & 12.179 (0.017) \\
56076.8 & 12.141 (0.017) & 12.172 (0.011) & 12.108 (0.011) & 12.110 (0.022) \\
56077.8 & 12.205 (0.011) & 12.150 (0.009) & 12.105 (0.009) & 12.061 (0.017) \\
56079.9 & 12.132 (0.008) & 12.018 (0.010) & 11.994 (0.008) & 12.032 (0.015) \\
56080.8 & 12.150 (0.013) & 11.995 (0.012) & 11.954 (0.009) & 12.058 (0.018) \\
56081.8 & 12.097 (0.017) & 11.951 (0.015) & 11.950 (0.012) & 12.055 (0.022) \\
56084.9 & 12.139 (0.010) & 11.904 (0.013) & 11.888 (0.012) & 12.097 (0.023) \\
56089.9 & 12.439 (0.016) & 12.055 (0.023) & 12.042 (0.024) & 12.352 (0.030) \\
56092.9 & 12.617 (0.018) & 12.258 (0.010) & 12.296 (0.008) & 12.540 (0.019) \\
56093.9 & 12.726 (0.018) & 12.325 (0.021) & 12.354 (0.012) & 12.606 (0.032) \\
56094.9 & 12.799 (0.025) & 12.355 (0.020) & 12.441 (0.013) & 12.635 (0.034) \\
56095.9 & 12.918 (0.033) & 12.462 (0.021) & 12.502 (0.014) & 12.659 (0.031) \\
56097.9 & 13.143 (0.032) & 12.576 (0.022) & 12.550 (0.018) & 12.659 (0.032) \\
56098.9 & 13.251 (0.027) & 12.629 (0.022) & 12.611 (0.010) & 12.647 (0.028) \\
56099.9 & 13.389 (0.025) & 12.677 (0.021) & 12.644 (0.017) & 12.630 (0.031) \\
56101.9 & 13.565 (0.025) & 12.834 (0.020) & 12.666 (0.010) & 12.607 (0.027) \\
56102.9 & 13.756 (0.029) & 12.884 (0.023) & 12.640 (0.018) & 12.571 (0.031) \\
56104.9 & 13.992 (0.032) & 12.951 (0.026) & 12.728 (0.013) & 12.568 (0.032) \\
\hline
\hline
\end{tabular}
\end{table*}

\begin{table*}
\caption{Local $BVRI$ standard stars in the vicinity of SN~2012ht}
\label{tab-12ht-std}
\centering
\tiny
\begin{tabular}{cccccccccc}
\hline
\hline
R.A. & Dec. & B & $\sigma_B$ & V & $\sigma_V$ & R & $\sigma_R$ & I & $\sigma_I$ \\
\hline
10:53:01.648 & +16:56:03.06 & 14.886 &  0.009 & 14.225 &  0.011 & 13.787 &  0.014 & 13.405 &  0.007 \\
10:54:06.561 & +16:45:53.13 & 15.589 &  0.005 & 14.944 &  0.006 & 14.697 &  0.005 & 14.505 &  0.009 \\
10:53:39.512 & +16:49:17.04 & 15.914 &  0.005 & 15.300 &  0.006 & 14.991 &  0.005 & 14.575 &  0.006 \\
10:54:01.309 & +16:50:07.90 & 15.759 &  0.005 & 15.149 &  0.006 & 14.841 &  0.005 & 14.426 &  0.006 \\
10:52:54.642 & +16:59:01.96 & 15.253 &  0.010 & 14.524 &  0.013 & 13.995 &  0.016 & 13.555 &  0.008 \\
10:53:05.252 & +16:50:56.91 & 15.838 &  0.005 & 15.149 &  0.006 & 14.799 &  0.005 & 14.354 &  0.006 \\
10:54:01.439 & +16:51:57.31 & 16.054 &  0.005 & 15.160 &  0.006 & 14.686 &  0.005 & 14.107 &  0.005 \\
10:52:57.121 & +16:54:56.27 & 15.751 &  0.006 & 14.975 &  0.007 & 14.497 &  0.007 & 13.958 &  0.007 \\
\hline
\hline
\end{tabular}
\end{table*}

\begin{table*}
\caption{Photometry of SN~2012ht. The $BVRI$ data are given in Vega-magnitudes, while the
$g'r'i'z'$ data are in AB-magnitudes.}
\label{tab-12ht-phot}
\centering
\tiny
\begin{tabular}{ccccccccc}
\hline
\hline
MJD & B & $\sigma_B$ & V & $\sigma_V$ & R & $\sigma_R$ & I & $\sigma_I$ \\
\hline
56066.8 & 14.995 &  0.031 & 14.679 &  0.025 & 14.593 &  0.023 & 14.420 &  0.048 \\
56067.9 & 14.493 &  0.014 & 14.221 &  0.010 & 14.133 &  0.011 & 13.974 &  0.018 \\
56069.8 & 13.629 &  0.014 & 13.490 &  0.010 & 13.375 &  0.009 & 13.229 &  0.018 \\
56072.9 & 12.765 &  0.008 & 12.674 &  0.010 & 12.575 &  0.008 & 12.462 &  0.017 \\
56073.8 & 12.646 &  0.015 & 12.475 &  0.016 & 12.481 &  0.012 & 12.360 &  0.019 \\
56075.8 & 12.452 &  0.012 & 12.175 &  0.012 & 12.233 &  0.007 & 12.179 &  0.017 \\
56076.8 & 12.141 &  0.017 & 12.172 &  0.011 & 12.108 &  0.011 & 12.110 &  0.022 \\
56077.8 & 12.205 &  0.011 & 12.150 &  0.009 & 12.105 &  0.009 & 12.061 &  0.017 \\
56079.9 & 12.132 &  0.008 & 12.018 &  0.010 & 11.994 &  0.008 & 12.032 &  0.015 \\
56080.8 & 12.150 &  0.013 & 11.995 &  0.012 & 11.954 &  0.009 & 12.058 &  0.018 \\
56081.8 & 12.097 &  0.017 & 11.951 &  0.015 & 11.950 &  0.012 & 12.055 &  0.022 \\
56084.9 & 12.139 &  0.010 & 11.904 &  0.013 & 11.888 &  0.012 & 12.097 &  0.023 \\
56089.9 & 12.439 &  0.016 & 12.055 &  0.023 & 12.042 &  0.024 & 12.352 &  0.030 \\
56092.9 & 12.617 &  0.018 & 12.258 &  0.010 & 12.296 &  0.008 & 12.540 &  0.019 \\
56093.9 & 12.726 &  0.018 & 12.325 &  0.021 & 12.354 &  0.012 & 12.606 &  0.032 \\
56094.9 & 12.799 &  0.025 & 12.355 &  0.020 & 12.441 &  0.013 & 12.635 &  0.034 \\
56095.9 & 12.918 &  0.033 & 12.462 &  0.021 & 12.502 &  0.014 & 12.659 &  0.031 \\
56097.9 & 13.143 &  0.032 & 12.576 &  0.022 & 12.550 &  0.018 & 12.659 &  0.032 \\
56098.9 & 13.251 &  0.027 & 12.629 &  0.022 & 12.611 &  0.010 & 12.647 &  0.028 \\
56099.9 & 13.389 &  0.025 & 12.677 &  0.021 & 12.644 &  0.017 & 12.630 &  0.031 \\
56101.9 & 13.565 &  0.025 & 12.834 &  0.020 & 12.666 &  0.010 & 12.607 &  0.027 \\
56102.9 & 13.756 &  0.029 & 12.884 &  0.023 & 12.640 &  0.018 & 12.571 &  0.031 \\
56104.9 & 13.992 &  0.032 & 12.951 &  0.026 & 12.728 &  0.013 & 12.568 &  0.032 \\
\hline
MJD & g & $\sigma_g$ & r & $\sigma_r$ & i & $\sigma_i$ & z & $\sigma_z$ \\
\hline
56285.94 & 14.182 &  0.227 & 14.074 &  0.094 & 14.277 &  0.087 & -- & -- \\
56290.01 & 13.508 &  0.107 & 13.444 &  0.039 & 13.657 &  0.062 & 13.867 &  0.075 \\
56291.14 & 13.347 &  0.151 & 13.375 &  0.094 & 13.673 &  0.073 & 13.730 &  0.107 \\
56292.11 & 13.383 &  0.147 & 13.250 &  0.115 & 13.513 &  0.119 & 13.883 &  0.126 \\
56303.97 & 13.402 &  0.140 & 13.518 &  0.119 & 14.306 &  0.167 & 14.335 &  0.173 \\
56315.94 & 14.700 &  0.270 & 14.037 &  0.109 & 14.457 &  0.215 & 14.213 &  0.410 \\
56322.94 & 15.277 &  0.288 & 14.537 &  0.098 & 14.658 &  0.188 & 14.368 &  0.189 \\
56326.86 & 15.458 &  0.202 & 14.757 &  0.064 & 14.873 &  0.189 & 14.551 &  0.178 \\
56330.97 & 15.601 &  0.129 & 14.964 &  0.078 & 15.180 &  0.110 & 14.953 &  0.211 \\
56331.91 & 15.529 &  0.047 & 15.003 &  0.113 & 15.195 &  0.121 & 14.988 &  0.170 \\
56340.92 & 15.975 &  0.153 & 15.477 &  0.100 & 15.616 &  0.138 & 15.482 &  0.157 \\
56351.86 & 16.162 &  0.226 & 15.896 &  0.143 & 15.649 &  0.224 & 15.316 &  0.293 \\
56353.86 & 16.122 &  0.122 & 15.815 &  0.135 & 16.081 &  0.188 & 16.085 &  0.156 \\
56355.96 & 16.199 &  0.199 & 15.931 &  0.146 & 16.201 &  0.203 & 16.393 &  0.254 \\
56356.80 & 16.392 &  0.181 & 15.958 &  0.181 & 16.205 &  0.191 & 16.624 &  0.289 \\
56362.91 & 16.372 &  0.208 & 16.206 &  0.149 & 16.424 &  0.224 & 16.468 &  0.236 \\
56388.80 & 16.875 &  0.175 & 17.059 &  0.151 & 17.456 &  0.232 & 17.738 &  0.394 \\
56395.91 & 17.092 &  0.166 & 17.087 &  0.045 & 17.444 &  0.143 & 18.357 &  0.488 \\
56396.80 & 16.949 &  0.146 & 17.281 &  0.152 & 17.330 &  0.241 & 18.708 &  1.002 \\
56397.80 & 16.947 &  0.133 & 17.180 &  0.060 & 17.666 &  0.157 & 18.496 &  0.410 \\
56398.86 & 16.930 &  0.093 & 17.348 &  0.100 & 17.499 &  0.104 & 17.457 &  0.191 \\

\hline
\hline
\end{tabular}
\end{table*}

\begin{table*}
\caption{Local $BVRI$ standard stars in the vicinity of SN~2013dy}
\label{tab-13dy-std}
\centering
\tiny
\begin{tabular}{cccccccccc}
\hline
\hline
R.A. & Dec. & B & $\sigma_B$ & V & $\sigma_V$ & R & $\sigma_R$ & I & $\sigma_I$ \\
\hline
22:18:22.19 & +40:34:21.9 & 16.233 & 0.032 & 15.606 & 0.014 & 15.243 & 0.015 & 14.844 & 0.016 \\
22:18:16.96 & +40:34:55.9 & 15.539 & 0.033 & 14.927 & 0.011 & 14.584 & 0.015 & 14.216 & 0.016 \\
\hline
\hline
\end{tabular}
\end{table*}

\begin{table*}
\caption{$BVRI$ photometry of SN~2013dy, given in the Vega-system}
%while the $g'r'i'z'$ data are in AB-magnitudes.}
\label{tab-13dy-phot1}
\centering
\tiny
\begin{tabular}{ccccccccc}
\hline
\hline
MJD & B & $\sigma_B$ & V & $\sigma_V$ & R & $\sigma_R$ & I & $\sigma_I$ \\
\hline
56490.9 & 14.069 &  0.008 & 13.824 &  0.033 & 13.653 &  0.045 & 13.522 &  0.031 \\
56492.0 & 13.929 &  0.021 & 13.636 &  0.032 & 13.459 &  0.064 & 13.341 &  0.065 \\
56493.0 & 13.720 &  0.093 & 13.479 &  0.022 & 13.307 &  0.062 & 13.205 &  0.040 \\
56494.0 & 13.584 &  0.067 & 13.359 &  0.011 & 13.196 &  0.070 & 13.103 &  0.038 \\
56494.9 & 13.492 &  0.028 & 13.271 &  0.002 & 13.118 &  0.059 & 13.042 &  0.039 \\
56495.9 & 13.388 &  0.064 & 13.168 &  0.038 & 13.050 &  0.067 & 12.980 &  0.055 \\
56496.9 & 13.376 &  0.052 & 13.118 &  0.041 & 12.998 &  0.088 & 12.955 &  0.038 \\
56497.9 & 13.325 &  0.045 & 13.036 &  0.066 & 12.954 &  0.057 & 12.950 &  0.030 \\
56498.8 & 13.282 &  0.027 & 13.028 &  0.022 & 12.934 &  0.051 & 12.957 &  0.042 \\
56499.8 & 13.274 &  0.076 & 13.043 &  0.002 & 12.902 &  0.063 & 12.970 &  0.058 \\
56500.8 & 13.316 &  0.002 & 12.981 &  0.021 & 12.913 &  0.068 & 12.983 &  0.043 \\
56501.8 & 13.255 &  0.002 & 12.957 &  0.034 & 12.874 &  0.063 & 13.010 &  0.048 \\
56505.8 & 13.415 &  0.010 & 13.006 &  0.040 & 12.939 &  0.059 & 13.135 &  0.055 \\
56506.9 & 13.427 &  0.030 & 13.044 &  0.030 & 12.960 &  0.065 & 13.203 &  0.026 \\
56507.9 & 13.489 &  0.044 & 13.070 &  0.048 & 13.004 &  0.075 & 13.250 &  0.049 \\
56509.8 & 13.610 &  0.064 & 13.153 &  0.019 & 13.112 &  0.061 & 13.369 &  0.059 \\
56511.8 & 13.748 &  0.074 & 13.268 &  0.029 & 13.248 &  0.069 & 13.499 &  0.060 \\
56512.8 & 13.827 &  0.068 & 13.335 &  0.033 & 13.332 &  0.054 & 13.569 &  0.034 \\
56520.9 & 14.717 &  0.060 & 13.801 &  0.012 & 13.622 &  0.051 & 13.614 &  0.045 \\
56521.9 & 14.819 &  0.039 & 13.849 &  0.058 & 13.652 &  0.055 & 13.588 &  0.054 \\
56534.8 & 15.934 &  0.001 & 14.382 &  0.045 & 13.972 &  0.064 & 13.528 &  0.030 \\
56536.8 & 16.040 &  0.013 & 14.518 &  0.031 & 14.089 &  0.043 & 13.619 &  0.048 \\
56538.8 & 16.127 &  0.040 & 14.624 &  0.016 & 14.198 &  0.073 & 13.750 &  0.028 \\
56539.8 & 16.166 &  0.043 & 14.677 &  0.007 & 14.265 &  0.053 & 13.808 &  0.027 \\
56541.8 & 16.215 &  0.024 & 14.782 &  0.044 & 14.389 &  0.065 & 13.928 &  0.053 \\
56542.8 & 16.271 &  0.047 & 14.843 &  0.018 & 14.433 &  0.068 & 13.998 &  0.046 \\
56554.9 & 16.465 &  0.038 & 15.236 &  0.017 & 14.898 &  0.052 & 14.621 &  0.059 \\
56563.0 & 16.568 &  0.022 & 15.473 &  0.036 & 15.167 &  0.046 & 14.980 &  0.044 \\
56566.8 & 16.598 &  0.035 & 15.559 &  0.025 & 15.288 &  0.062 & 15.160 &  0.035 \\
56573.8 & 16.701 &  0.054 & 15.760 &  0.023 & 15.513 &  0.053 & 15.424 &  0.043 \\
56577.8 & 16.730 &  0.042 & 15.866 &  0.027 & 15.625 &  0.059 & 15.588 &  0.014 \\
56590.9 & 16.905 &  0.061 & 16.182 &  0.028 & 16.021 &  0.067 & 16.073 &  0.046 \\
56591.9 & 16.912 &  0.018 & 16.203 &  0.043 & 16.052 &  0.050 & 16.099 &  0.027 \\
56596.7 & 16.973 &  0.042 & 16.328 &  0.012 & 16.200 &  0.035 & 16.294 &  0.030 \\
56603.9 & 17.017 &  0.054 & 16.584 &  0.026 & 16.424 &  0.064 & 16.540 &  0.066 \\
\hline
\hline
\end{tabular}
\end{table*}

\begin{table*}
\caption{$g'r'i'z'$ photometry of SN~2013dy, given as AB-magnitudes}
\label{tab-13dy-phot2}
\centering
\tiny
\begin{tabular}{ccccccccc}
\hline
\hline
MJD & g & $\sigma_g$ & r & $\sigma_r$ & i & $\sigma_i$ & z & $\sigma_z$ \\
\hline
56490.92 & 13.901 &  0.079 & 13.689 &  0.060 & 13.789 &  0.050 & 13.901 &  0.131 \\
56491.90 & 13.705 &  0.110 & 13.481 &  0.057 & 13.557 &  0.078 & 13.607 &  0.121 \\
56492.91 & 13.623 &  0.101 & 13.332 &  0.068 & 13.365 &  0.086 & 13.527 &  0.115 \\
56493.91 & 13.534 &  0.124 & 13.198 &  0.041 & 13.320 &  0.047 & 13.463 &  0.074 \\
56494.91 & 13.462 &  0.115 & 13.174 &  0.051 & 13.313 &  0.063 & 13.440 &  0.083 \\
56495.88 & 13.377 &  0.130 & 13.100 &  0.056 & 13.271 &  0.068 & 13.364 &  0.079 \\
56496.84 & 13.176 &  0.103 & 13.036 &  0.094 & 13.257 &  0.092 & 13.473 &  0.203 \\
56497.85 & 13.349 &  0.134 & 13.022 &  0.050 & 13.272 &  0.065 & 13.336 &  0.121 \\
56498.89 & 13.170 &  0.104 & 12.924 &  0.079 & 13.201 &  0.074 & 13.237 &  0.080 \\
56502.90 & 13.036 &  0.061 & 12.980 &  0.063 & 13.237 &  0.119 & 13.335 &  0.138 \\
56503.93 & 13.017 &  0.070 & 12.888 &  0.032 & 13.366 &  0.045 & 13.322 &  0.108 \\
56504.88 & 13.120 &  0.082 & 12.882 &  0.036 & 13.373 &  0.059 & 13.330 &  0.099 \\
56505.88 & 12.998 &  0.058 & 12.895 &  0.041 & 13.449 &  0.044 & 13.334 &  0.088 \\
56508.88 & 13.160 &  0.068 & 13.005 &  0.038 & 13.558 &  0.079 & 13.605 &  0.108 \\
56509.84 & 13.217 &  0.071 & 13.112 &  0.048 & 13.587 &  0.077 & 13.579 &  0.116 \\
56510.86 & 13.209 &  0.048 & 13.175 &  0.033 & 13.726 &  0.041 & 13.600 &  0.155 \\
56511.84 & 13.426 &  0.083 & 13.222 &  0.053 & 13.745 &  0.064 & 13.656 &  0.134 \\
56512.85 & 13.394 &  0.064 & 13.315 &  0.031 & 13.872 &  0.054 & 13.842 &  0.136 \\
56513.85 & 13.622 &  0.085 & 13.416 &  0.051 & 13.938 &  0.102 & 13.670 &  0.085 \\
56515.86 & 13.520 &  0.050 & 13.496 &  0.032 & 14.060 &  0.050 & 13.651 &  0.083 \\
56519.83 & 14.033 &  0.108 & 13.608 &  0.045 & 14.033 &  0.074 & 13.647 &  0.096 \\
56520.83 & 14.096 &  0.097 & 13.617 &  0.053 & 14.086 &  0.065 & 13.736 &  0.143 \\
56521.82 & 14.133 &  0.091 & 13.660 &  0.055 & 14.036 &  0.064 & 13.644 &  0.124 \\
56526.82 & 14.525 &  0.159 & 13.803 &  0.088 & 13.916 &  0.097 & 13.582 &  0.087 \\
56530.87 & 14.663 &  0.067 & 13.826 &  0.054 & 13.913 &  0.053 & 13.580 &  0.125 \\
56534.87 & 14.812 &  0.061 & 13.959 &  0.029 & 13.977 &  0.035 & 13.640 &  0.100 \\
56535.79 & 14.929 &  0.077 & 14.060 &  0.046 & 14.121 &  0.096 & 13.792 &  0.215 \\
56538.82 & 15.068 &  0.053 & 14.214 &  0.044 & 14.200 &  0.061 & 13.836 &  0.112 \\
56539.81 & 15.108 &  0.067 & 14.252 &  0.033 & 14.261 &  0.043 & 13.921 &  0.126 \\
56541.79 & 15.152 &  0.051 & 14.348 &  0.035 & 14.371 &  0.059 & 13.991 &  0.126 \\
56542.83 & 15.225 &  0.057 & 14.380 &  0.031 & 14.412 &  0.064 & 14.071 &  0.126 \\
56543.83 & 15.377 &  0.118 & 14.487 &  0.051 & 14.463 &  0.067 & 14.187 &  0.098 \\
56551.76 & 15.989 &  0.190 & 14.911 &  0.082 & 15.037 &  0.113 & 14.768 &  0.153 \\
56552.80 & 15.496 &  0.103 & 14.731 &  0.054 & 14.866 &  0.077 & 14.616 &  0.123 \\
56554.83 & 15.782 &  0.162 & 14.813 &  0.043 & 14.900 &  0.052 & 14.880 &  0.101 \\
56557.79 & 15.536 &  0.100 & 14.899 &  0.062 & 14.908 &  0.075 & 14.920 &  0.150 \\
56558.90 & 15.572 &  0.085 & 14.879 &  0.036 & 15.002 &  0.053 & 14.842 &  0.099 \\
56559.77 & 15.496 &  0.066 & 14.939 &  0.033 & 15.049 &  0.046 & 15.022 &  0.142 \\
56560.76 & 15.517 &  0.078 & 14.978 &  0.052 & 15.071 &  0.068 & 15.149 &  0.207 \\
56568.79 & 15.570 &  0.068 & 15.117 &  0.042 & 15.269 &  0.063 & 15.257 &  0.082 \\
56569.77 & 15.636 &  0.066 & 15.188 &  0.058 & 15.240 &  0.064 & 15.241 &  0.135 \\
56575.91 & 15.752 &  0.074 & 15.262 &  0.049 & 15.444 &  0.067 & 15.529 &  0.129 \\
56578.73 & 15.815 &  0.095 & 15.325 &  0.046 & 15.479 &  0.055 & 15.537 &  0.116 \\
56582.90 & 16.002 &  0.110 & 15.504 &  0.062 & 15.520 &  0.068 & 15.640 &  0.119 \\
56584.74 & 15.941 &  0.132 & 15.432 &  0.055 & 15.952 &  0.131 & 15.511 &  0.131 \\
56586.78 & 15.976 &  0.102 & 15.548 &  0.058 & 15.726 &  0.074 & 15.618 &  0.108 \\
56588.81 & 15.836 &  0.091 & 15.538 &  0.058 & 15.659 &  0.060 & 15.863 &  0.136 \\
56590.80 & 15.880 &  0.081 & 15.650 &  0.069 & 15.705 &  0.061 & 15.781 &  0.139 \\
56591.77 & 15.842 &  0.061 & 15.533 &  0.038 & 15.645 &  0.059 & 15.759 &  0.109 \\
56594.73 & 15.907 &  0.048 & 15.652 &  0.050 & 15.743 &  0.063 & 16.255 &  0.201 \\
56595.98 & 16.008 &  0.092 & 15.909 &  0.129 & 15.789 &  0.102 & 16.586 &  0.356 \\
56597.97 & 15.893 &  0.134 & 15.911 &  0.126 & 15.958 &  0.182 & 16.909 &  0.590 \\
56603.75 & 15.991 &  0.052 & 15.709 &  0.037 & 15.882 &  0.052 & 15.871 &  0.103 \\
56627.82 & 16.224 &  0.072 & 16.012 &  0.058 & 16.025 &  0.076 & 16.152 &  0.180 \\
56628.82 & 16.319 &  0.085 & 16.127 &  0.089 & 16.120 &  0.084 & 15.890 &  0.149 \\
\hline
\hline
\end{tabular}
\end{table*}

\begin{table*}
\caption{Local $BVRI$ standard stars in the vicinity of SN~2014J}
\label{tab-14J-std}
\centering
\tiny
\begin{tabular}{cccccccccc}
\hline
\hline
R.A. & Dec. & B & $\sigma_B$ & V & $\sigma_V$ & R & $\sigma_R$ & I & $\sigma_I$ \\
\hline
09:56:37.96 & +69:41:18.9 & 15.024 & 0.012 & 14.277 & 0.010 & 13.855 & 0.007 & 13.396 & 0.010 \\
09:56:32.99 & +69:39:17.9 & 13.983 & 0.008 & 13.504 & 0.007 & 13.209 & 0.005 & 12.816 & 0.008 \\
\hline
\hline
\end{tabular}
\end{table*}

\begin{table*}
\caption{Photometry of SN~2014J. The $BVRI$ data are given in Vega-magnitudes, while the $g'r'i'z'$ data are in AB-magnitudes.}
\label{tab-14J-phot}
\centering
\tiny
\begin{tabular}{ccccccccc}
\hline
\hline
MJD & B & $\sigma_B$ & V & $\sigma_V$ & R & $\sigma_R$ & I & $\sigma_I$ \\
\hline
56684.0 & 12.142 &  0.040 & 10.921 &  0.018 & 10.282 &  0.003 &  9.802 &  0.031 \\
56688.8 & 11.846 &  0.068 & 10.669 &  0.015 & 10.093 &  0.028 &  9.697 &  0.002 \\
56691.0 & 11.971 &  0.117 & 10.622 &  0.025 & 10.040 &  0.008 &  9.827 &  0.004 \\
56692.0 & 11.931 &  0.081 & 10.588 &  0.037 & 10.071 &  0.015 &  9.858 &  0.020 \\
56693.0 & 11.965 &  0.032 & 10.584 &  0.031 & 10.078 &  0.027 &  9.819 &  0.011 \\
56703.0 & 12.818 &  0.020 & 11.136 &  0.022 & 10.655 &  0.013 & 10.367 &  0.005 \\
56706.2 & 13.211 &  0.071 & 11.345 &  0.041 & 10.835 &  0.018 & 10.348 &  0.027 \\
56709,0 & 13.538 &  0.085 & 11.362 &  0.071 & 10.770 &  0.037 & 10.280 &  0.078 \\
56712.0 & 13.745 &  0.020 & 11.458 &  0.027 & 10.789 &  0.007 & 10.122 &  0.004 \\
56717.8 & 14.521 &  0.001 & 11.726 &  0.002 & 10.995 &  0.004 & 10.184 &  0.012 \\
56724.9 & 14.822 &  0.024 & 12.179 &  0.021 & 11.326 &  0.009 & 10.425 &  0.010 \\
56726.0 & 14.944 &  0.035 & 12.152 &  0.026 & 11.371 &  0.014 & 10.480 &  0.015 \\
56727.0 & 15.009 &  0.040 & 12.250 &  0.008 & 11.457 &  0.003 & 10.545 &  0.019 \\
56728.0 & 14.969 &  0.006 & 12.343 &  0.040 & 11.537 &  0.018 & 10.632 &  0.020 \\
56729.0 & 15.060 &  0.006 & 12.367 &  0.041 & 11.550 &  0.001 & 10.687 &  0.030 \\
56730.9 & 15.003 &  0.002 & 12.438 &  0.019 & 11.650 &  0.023 & 10.643 &  0.008 \\
56736.9 & 15.042 &  0.004 & 12.659 &  0.006 & 11.904 &  0.005 & 11.148 &  0.015 \\
56739.0 & 15.124 &  0.017 & 12.714 &  0.016 & 11.958 &  0.043 & 11.034 &  0.024 \\
56741.9 & 15.169 &  0.018 & 12.792 &  0.019 & 12.035 &  0.002 & 11.238 &  0.005 \\
56742.8 & 15.156 &  0.001 & 12.809 &  0.001 & 12.065 &  0.001 & 11.371 &  0.001 \\
56744.8 & 15.142 &  0.026 & 12.855 &  0.002 & 11.920 &  0.014 & 11.370 &  0.012 \\
56746.0 & 15.198 &  0.017 & 12.916 &  0.022 & 12.092 &  0.005 & 11.411 &  0.011 \\
56746.9 & 15.200 &  0.001 & 12.938 &  0.007 & 12.060 &  0.007 & 11.480 &  0.009 \\
56751.8 & 15.223 &  0.019 & 13.045 &  0.018 & 12.318 &  0.008 & 11.711 &  0.025 \\
56754.8 & 15.248 &  0.030 & 13.124 &  0.008 & 12.420 &  0.007 & 11.790 &  0.006 \\
56774.8 & 15.215 &  0.029 & 13.602 &  0.004 & 12.945 &  0.007 & 12.480 &  0.013 \\
56777.9 & 15.407 &  0.008 & 13.682 &  0.021 & 13.062 &  0.004 & 12.608 &  0.035 \\
56785.8 & 15.405 &  0.043 & 13.829 &  0.002 & 13.199 &  0.010 & 12.731 &  0.008 \\
56798.9 & 15.571 &  0.014 & 14.108 &  0.020 & 13.545 &  0.007 & 13.078 &  0.025 \\
\hline
MJD & g & $\sigma_g$ & r & $\sigma_r$ & i & $\sigma_i$ & z & $\sigma_z$ \\
\hline
56680.05 & 12.295 &  0.030 & 10.941 &  0.056 & 10.836 &  0.023 & 10.415 &  0.034 \\
56682.97 & 11.733 &  0.033 & 10.554 &  0.063 & 10.397 &  0.013 & 10.072 &  0.037 \\
56683.97 & 11.582 &  0.018 & 10.504 &  0.081 & 10.281 &  0.033 &  9.872 &  0.058 \\
56691.07 & 11.267 &  0.014 & 10.213 &  0.012 & 10.311 &  0.019 &  9.977 &  0.029 \\
56692.10 & 11.278 &  0.016 & 10.199 &  0.007 & 10.340 &  0.016 & 10.018 &  0.032 \\
56692.97 & 11.284 &  0.011 & 10.218 &  0.012 & 10.354 &  0.016 & 10.044 &  0.038 \\
56698.12 & 11.414 &  0.016 & 10.435 &  0.004 & 10.610 &  0.026 & 10.172 &  0.024 \\
56700.98 & 11.540 &  0.017 & 10.694 &  0.020 & 10.967 &  0.046 & 10.269 &  0.033 \\
56701.79 & 11.627 &  0.010 & 10.719 &  0.010 & 10.908 &  0.031 & 10.254 &  0.033 \\
56702.97 & 11.687 &  0.012 & 10.787 &  0.011 & 10.970 &  0.028 & 10.267 &  0.039 \\
56703.76 & 11.718 &  0.005 & 10.851 &  0.006 & 10.970 &  0.023 & 10.259 &  0.052 \\
56713.90 & 12.509 &  0.017 & 10.958 &  0.010 & 10.821 &  0.032 & 10.075 &  0.037 \\
56714.84 & 12.518 &  0.015 & 10.990 &  0.014 & 10.749 &  0.021 & 10.104 &  0.034 \\
56715.77 & 12.572 &  0.029 & 10.987 &  0.017 & 10.734 &  0.019 & 10.091 &  0.062 \\
56717.85 & 12.697 &  0.012 & 11.069 &  0.011 & 10.815 &  0.030 & 10.124 &  0.038 \\
56719.91 & 12.879 &  0.011 & 11.134 &  0.010 & 10.878 &  0.023 & 10.116 &  0.030 \\
56724.97 & 13.230 &  0.009 & 11.489 &  0.012 & 11.133 &  0.026 & 10.336 &  0.035 \\
56725.86 & 13.225 &  0.021 & 11.581 &  0.010 & 11.182 &  0.026 & 10.443 &  0.035 \\
56727.93 & 13.375 &  0.014 & 11.650 &  0.016 & 11.336 &  0.030 & 10.509 &  0.036 \\
56728.93 & 13.467 &  0.020 & 11.747 &  0.012 & 11.424 &  0.029 & 10.584 &  0.034 \\
56729.95 & 13.473 &  0.027 & 11.765 &  0.016 & 11.489 &  0.035 & 10.620 &  0.030 \\
56730.80 & 13.400 &  0.022 & 11.807 &  0.012 & 11.471 &  0.030 & 10.752 &  0.034 \\
56733.93 & 13.552 &  0.016 & 11.961 &  0.017 & 11.614 &  0.032 & 10.938 &  0.029 \\
56734.75 & 13.457 &  0.047 & 11.899 &  0.057 & 11.565 &  0.035 & 11.144 &  0.053 \\
56737.85 & 13.627 &  0.013 & 12.047 &  0.018 & 11.760 &  0.033 & 11.074 &  0.040 \\
56745.79 & 13.815 &  0.011 & 12.297 &  0.019 & 12.063 &  0.039 & 11.482 &  0.040 \\
56746.92 & 13.748 &  0.104 & 12.300 &  0.035 & 12.112 &  0.051 & 11.655 &  0.119 \\
56754.82 & 13.980 &  0.016 & 12.558 &  0.022 & 12.293 &  0.040 & 11.866 &  0.049 \\
56768.80 & 14.227 &  0.020 & 12.910 &  0.030 & 12.682 &  0.047 & 12.493 &  0.052 \\
56769.88 & 14.236 &  0.018 & 12.965 &  0.029 & 12.715 &  0.044 & 12.390 &  0.058 \\
56773.86 & 14.314 &  0.018 & 13.079 &  0.028 & 12.852 &  0.046 & 12.551 &  0.058 \\
56775.88 & 14.255 &  0.028 & 13.095 &  0.034 & 12.896 &  0.054 & 12.976 &  0.114 \\
56782.84 & 14.382 &  0.024 & 13.294 &  0.029 & 13.037 &  0.052 & 12.750 &  0.058 \\
56783.84 & 14.464 &  0.018 & 13.346 &  0.028 & 13.060 &  0.056 & 12.849 &  0.069 \\
56785.81 & 14.512 &  0.019 & 13.361 &  0.032 & 13.098 &  0.052 & 12.828 &  0.062 \\
56787.86 & 14.547 &  0.022 & 13.384 &  0.032 & 13.114 &  0.055 & 12.887 &  0.069 \\
56796.91 & 14.654 &  0.029 & 13.628 &  0.049 & 13.388 &  0.068 & 13.056 &  0.081 \\
56797.88 & 14.672 &  0.025 & 13.604 &  0.049 & 13.312 &  0.068 & 13.033 &  0.085 \\
56798.84 & 14.670 &  0.029 & 13.603 &  0.043 & 13.245 &  0.057 & 12.918 &  0.068 \\
56799.89 & 14.686 &  0.022 & 13.633 &  0.037 & 13.305 &  0.060 & 13.097 &  0.076 \\
56805.87 & 14.834 &  0.028 & 13.747 &  0.054 & 13.362 &  0.077 & 13.156 &  0.084 \\
56811.86 & 14.898 &  0.035 & 13.893 &  0.052 & 13.530 &  0.072 & 13.187 &  0.088 \\
\hline
\hline
\end{tabular}
\end{table*}

\end{appendix}

\end{document}